\documentclass[8.5pt,twoside,a4paper]{article}
\usepackage{times,mathptm}
\usepackage{sectsty}
\usepackage{balance} 
\usepackage{amsmath,amssymb,amsthm}             % AMS Math
\usepackage{booktabs}
\usepackage{sidecap}
\usepackage{threeparttable}
\usepackage{color}

\usepackage[T1]{fontenc} % Modern font encoding
\usepackage{float}       % For creating charts, graphs and schemes
\usepackage{helvet}      % Helvetica font for sans serif
\usepackage{mathptmx}    % Times font ("Word-like")
\usepackage{rsc}         % Loads natbib, etc.
\usepackage{graphicx} %eps figures can be used instead
\usepackage{lastpage}
\usepackage[format=plain,justification=raggedright,singlelinecheck=false,font=small,labelfont=bf,labelsep=space]{caption} 
\usepackage{fancyhdr}
\newcommand{\GILDAS}{\texttt{GILDAS}}
\newcommand{\CLASS}{\texttt{CLASS}}

\newcommand{\ie}{\emph{i.e.}}
\newcommand{\eg}{e.g.}
\newcommand{\emm}[1]{\ensuremath{#1}}   % Ensures math mode.
\newcommand{\emr}[1]{\emm{\mathrm{#1}}} % Uses math roman fonts.
\newcommand{\Tkin}{\emm{T_\emr{kin}}}

\newcommand{\Td}{\emm{T_\emr{dust}}}

\newcommand{\dcop}{\emr{DCO^{+}}}                  % DCO+
\newcommand{\h}{\emr{H}}                          % H
\newcommand{\x}{\emr{X}}                          % H
\newcommand{\hhco}{\emr{H_2CO}}            % H2CO
\newcommand{\ohhco}{\emr{o-H_2CO}}         % o-H2CO
\newcommand{\phhco}{\emr{p-H_2CO}}         % p-H2CO
\newcommand{\chhhoh}{\emr{CH_3OH}}         % CH3OH
\newcommand{\chhhohE}{\emr{CH_3OH-E}}         % CH3OH-E
\newcommand{\chhhohA}{\emr{CH_3OH-A}}         % CH3OH-A
\newcommand{\ddco}{\emr{D_2CO}}            % D2CO
\newcommand{\hh}{\emr{H_2}}                       % H2
\newcommand{\hho}{\emr{H_2O}}              % H2O
\newcommand{\cp}{\emr{C^+}}                      % C+
\newcommand{\cfp}{\emr{CF^+}}                    % CF+
\newcommand{\hcooh}{\emr{HCOOH}}         % HCOOH
\newcommand{\thcooh}{\emr{t-HCOOH}}         % t-HCOOH
\newcommand{\chhhcch}{\emr{CH_3CCH}}         % CH3CCH
\newcommand{\chhhcchE}{\emr{CH_3CCH-E}}         % CH3C2H-E
\newcommand{\chhhcchA}{\emr{CH_3CCH-A}}         % CH3C2H-A
\newcommand{\chhco}{\emr{CH_2CO}}         % CH2CO
\newcommand{\ochhco}{\emr{o-CH_2CO}}         % o-CH2CO
\newcommand{\pchhco}{\emr{p-CH_2CO}}         % p-CH2CO
\newcommand{\chhhcho}{\emr{CH_3CHO}}         % CH3CHO
\newcommand{\chhhchoE}{\emr{CH_3CHO-E}}         % CH3CHO-E
\newcommand{\chhhchoA}{\emr{CH_3CHO-A}}         % CH3CHO-A

\newcommand{\hcop}{\emr{HCO^+}}                    % HCO+
\newcommand{\cch}{\emr{C_2H}}                    % CCH
\newcommand{\ccch}{\emr{C_3H}}                    % C3H
\newcommand{\ccchh}{\emr{C_3H_2}}                    % C3H2
\newcommand{\ccchp}{\emr{C_3H^+}}                    % C3H+
\newcommand{\hcccn}{\emr{HC_3N}}                   % HC3N 
\newcommand{\chhhcn}{\emr{CH_3CN}}                   % CH3CN
\newcommand{\chhhnc}{\emr{CH_3NC}}                   % CH3NC
\newcommand{\hthcop}{\emr{H^{13}CO^+}}                   % H13CO+
\newcommand{\ctfs}{\emr{C^{34}S}}                   % C34S
\newcommand{\ceio}{\emr{C^{18}O}}                   % C18S
\newcommand{\unit}[1]{\emm{\, \emr{#1}}}
\newcommand{\mm}{\unit{mm}}
\renewcommand{\deg}{\emm{^\circ}}

\newcommand{\pscm}{~\rm{cm}^{-2}}
\newcommand{\ps}{~\rm{s}^{-1}}
\newcommand{\kms}{\emr{\,km\,s^{-1}}}

\newcommand{\mKkms}{~\rm{mK\,km\,s}^{-1}}
\newcommand{\Tsys}{\emm{T_\emr{sys}}}
\newcommand{\Tas}{\emm{T_\emr{A}^*}}
\newcommand{\Tmb}{\emm{T_\emr{mb}}}
\newcommand{\Beff}{\emm{B_\emr{eff}}}
\newcommand{\Feff}{\emm{F_\emr{eff}}}
\begin{document}

\newcommand{\TabObsMaps}{%
  \begin{table*}[t!]
    \footnotesize{
    \centering
    \caption{Observation parameters for the maps shown in
      Figs.~\ref{fig:maps}. The projection center of the maps is
      $\alpha_{2000} = 05^h40^m54.27^s$, $\delta_{2000} = -02\deg 28'
      00''$.}
    \begin{threeparttable}
      {\footnotesize
        \begin{tabular}{ccrccccccl}
          \toprule
          Molecule & Transition & Frequency  & Instrument & Beam &
          PA & Int. Time & \Tsys{} & Noise & Ref.\\
          & & GHz & & arcsec & $^{\circ}$ & hours & K (\Tas{}) & K (\Tmb{}) &\\
      \midrule
      \multicolumn{2}{c}{Continuum at 1.2\mm} & 250.000000 & 30m/MAMBO  & 11.7 &  --  & -- &  -- & -- & \citet{hily-blant2005}\\
      CCH & $1,3/2(2)-0,1/2(1)$ & 87.316898 & PdBI/C\&D & $7.2\times5.0$ & 54 & & &  & \citet{pety2005} \\
      \chhhcho{} & $5_{15}-4_{14},6_{16}-5_{15}$ &  93.5,112.2& 30m/EMIR & 30.0 &0 & & & & This work\\
      %\chhhchoE{} & $5_{15}-4_{14}$ &  93.595235 & 30m/EMIR & 30.0 &0 & & & & This work\\
      %\chhhchoA{} & $6_{16}-5_{15}$ & 112.248716 & 30m/EMIR & 30.0 &0 & & & & This work\\
      %\chhhchoE{} & $6_{16}-5_{15}$ & 112.254508 & 30m/EMIR & 30.0 &0 & & & & This work\\
      HCO & $1_{01} 3/2,2-0_{00} 1/2,1$ & 86.670760 & 30m/AB100 & 29.9 & 0 & 2.6/5.0$^{a}$ & 133 & 63 & \citet{gerin2009}\\   
      \cfp{} & $1-0$ & 102.587533 & 30m/EMIR & 25.4 & 0 & 2.5 & 88 & 0.13 & \citet{guzman2012a}\\
      \dcop{} & $3-2$ & 216.112582 & 30m/HERA & 11.4 & 0 & 1.5/2.0$^{a}$ & 230 & 0.10 & \citet{pety2007}\\
      $\phhco$ & $2_{02}-1_{01}$ & 145.602949 & 30m/EMIR & 17.8 & 0 & 7.4/12.9$^a$ & 208 & 0.17 & \citet{guzman2013}\\
      $\chhhohA$ & $3_{0}-2_{0}$ & 145.103152 & 30m/EMIR & 17.9 & 0 & 7.4/12.9$^a$ & 208 & 0.095 & \citet{guzman2013}\\
      \bottomrule
\end{tabular}}
\begin{tablenotes}[para,flushleft]
  $^{a}$ Two values are given for the integration time: the
  on-source time and the telescope time.\\
\end{tablenotes}    
    \end{threeparttable}
    \label{tab:obs:maps} }
  \end{table*}
  }

\newcommand{\TabDipole}{%
  \begin{table}[t!]
    \caption{Dipole moments of complex molecules}
    \begin{tabular}{lcr}\toprule
      Species & Dipole moment & Reference\\
      & (Debye)&\\ 
      \midrule
      \hcooh     & 1.4 & \citet{kim1962}\\
      \chhco{}   & 1.45 &\citet{hannay1946} \\
      \chhhcho{} & 2.7 & \citet{kleiner1996}\\
      \chhhcch{} & 0.78 & \citet{burrell1980} \\
      \bottomrule
    \end{tabular}
    \label{tab:dipole}
\end{table}
}

\newcommand{\TabAbundances}{%
\begin{table}[t!]
  \caption{Summary of abundances with respect to total hydrogen nuclei
    ($N_\mathrm{H} = N(\x)/(N(\h)+2~N(\hh))$) toward the PDR and
    dense core. The column densities of the total hydrogen nuclei are
    $N_\mathrm{H}=3.8\times10^{22} \pscm$ (PDR) and
    $N_\mathrm{H}=6.4\times10^{22} \pscm$ (dense core).} \centering
  \begin{threeparttable}
    \begin{tabular}{llccccc}\toprule
      Species & Beam & \multicolumn{2}{c}{PDR} &
      \multicolumn{2}{c}{Core} & Ref. \\
      &  ($''$) & Abundance & Offsets & Abundance & Offsets & \\
      \midrule
      %$N(\hh{})$ & 11             & $1.9\times10^{22}$ & (-5,0) & $2.9\times10^{22}$ & (20,22) & \cite{goicoechea2006,pety2005,gerin2009,goicoechea2009b}\\
      %\midrule
      %C\,{\sc i} &\\
      %O\,{\sc i} &\\
      %CO         &\\
      %$^{13}$CO  &\\
      \ceio{}    & $6.5\times4.3$ & $1.9\times10^{-7}$   & (-6,4)  & -& -& \citet{pety2005}\\
      \cch{}     & $7.2\times5.0$ & $1.4\times10^{-8}$   & (-6,4)  & -&- & \citet{pety2005}\\
      c-\ccch{}  & $28$           & $2.7\times10^{-10}$  & (-10,0) & -&-& \citet{teyssier2004}\\
      l-\ccch{}  & $28$           & $1.4\times10^{-10}$  & (-10,0) & -&-& \citet{teyssier2004}\\
      c-\ccchh{} & $6.1\times4.7$ & $1.1\times10^{-9}$   & (-6,4)  &- & -& \citet{pety2005}\\
      l-\ccchh{} & $27$           & $<4.6\times10^{-11}$ & (-10,0) &- &-& \citet{teyssier2004}\\
      C$_4$H     & $6.1\times4.7$ & $1.0\times10^{-9}$   & (-6,-4) &- &- & \citet{pety2005}\\
      C$_6$H     & $28$           & $2.2\times10^{-11}$  & (-6,4)  &- &- & \citet{agundez2008}\\
      CS         & $10$           & $2.0\times10^{-9}$   & (4,0)   & $2.9\times10^{-9}$ & (21,15) & \citet{goicoechea2006}\\
      \ctfs{}    & $16$           & $9.2\times10^{-11}$  & (4,0)   & $9.1\times10^{-11}$ & (21,15) & \citet{goicoechea2006}\\
      HCS$^+$    & $29$           & $1.7\times10^{-11}$  & (4,0)   & $1.2\times10^{-11}$ & (21,15) & \citet{goicoechea2006}\\
      HCO        & $6.7\times4.4$ & $8.4\times10^{-10}$  & (-5,0)  & $<8.0\times10^{-11}$ & (20,22) & \citet{gerin2009}\\
      \hcop{}    & $28$           & $9.0\times10^{-10}$  & (-5,0)  & $3.9\times10^{-9}$  & (20,22) &\citet{goicoechea2009b}\\
      \hthcop{}  & $6.8\times4.7$ & $1.5\times10^{-11}$  & (-5,0)  & $6.5\times10^{-11}$ & (20,22) &\citet{goicoechea2009b}\\
      HOC$^+$    & $28$           & $4.0\times10^{-12}$  & (-5,0)  & -                & (20,22) & \citet{goicoechea2009b}\\ 
      CO$^+$     & $10$           & $<5.0\times10^{-13}$ & (-5,0)  & -                & (20,22) & \citet{goicoechea2009b}\\
      \dcop{}    & $12$           & -                   & (-5,0)  & $8.0\times10^{-11}$ & (20,22) & \citet{pety2007}\\
      \cfp{}     & $25$           & $5.7\times10^{-10}$  & (-5,0)  & $<6.9\times10^{-11}$ & (20,22) & \citet{guzman2012a}\\
      \ccchp{}   & $27$           & $3.1\times10^{-11}$  & (-5,0)  & - & (20,22) & \citet{pety2012}\\
      \ohhco{}   & $6.1\times5.6$ & $1.9\times10^{-10}$  & (-5,0)  & $1.5\times10^{-10}$ & (20,22) & \citet{guzman2011}\\
      \phhco{}   & $6.1\times5.6$ & $9.5\times10^{-11}$  & (-5,0)  & $5.0\times10^{-11}$ & (20,22) & \citet{guzman2011}\\
      HDCO       & $18$           & -                   & (-5,0)  & $2.5\times10^{-11}$ & (20,22) & \citet{guzman2011}\\
      \ddco{}    & $24$           & -                   & (-5,0)  & $1.6\times10^{-11}$ & (20,22) & \citet{guzman2011}\\
      \chhhohE{} & $6.1\times5.6$ & $7.0\times10^{-11}$  & (-5,0)  & $1.0\times10^{-10}$ & (20,22) & \citet{guzman2013}\\
      \chhhohA{} & $6.1\times5.6$ & $5.3\times10^{-11}$  & (-5,0)  & $1.3\times10^{-10}$ & (20,22) & \citet{guzman2013}\\
      \chhhcn{}  & $27$           & $2.5\times10^{-10}$  & (-5,0)  & $7.9\times10^{-12}$ & (20,22) & \citet{gratier2013}\\
      \chhhnc{}  & $25$           & $4.1\times10^{-11}$  & (-5,0)  & $<7.8\times10^{-12}$ & (20,22) & \citet{gratier2013}\\
      \hcccn     & $30$           & $6.3\times10^{-12}$  & (-5,0)  & $7.9\times10^{-12}$ & (20,22) & \citet{gratier2013}\\
      \thcooh    & $29$           & $5.2\times10^{-11}$  & (-5,0)  & $1.4\times10^{-11}$ & (20,22) & This work\\
      \ochhco    & $30$           & $1.3\times10^{-10}$  & (-5,0)  & $4.2\times10^{-11}$ & (20,22) & This work\\
      \pchhco    & $26$           & $1.8\times10^{-11}$  & (-5,0)  & $7.3\times10^{-12}$ & (20,22) & This work\\
      \chhhchoE  & $27$           & $1.4\times10^{-11}$  & (-5,0)  & $3.9\times10^{-12}$ & (20,22) & This work\\
      \chhhchoA  & $27$           & $5.4\times10^{-11}$  & (-5,0)  & $2.0\times10^{-11}$ & (20,22) & This work\\
      \chhhcch   & $29$           & $4.4\times10^{-10}$  & (-5,0)  & $3.0\times10^{-10}$ & (20,22) & This work\\
      \bottomrule
    \end{tabular}
  \end{threeparttable}
  \label{tab:abundances}
\end{table}
}

\newcommand{\FigMaps}{%
\begin{figure*}[t!]
  \centering
  \includegraphics[width=\textwidth]{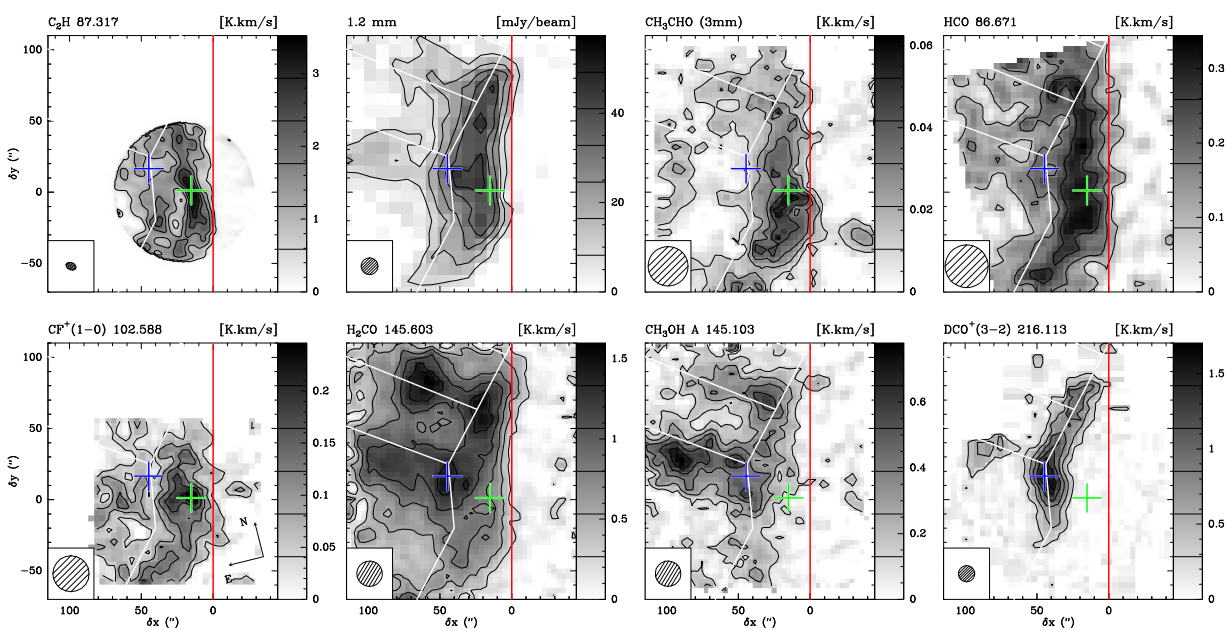} 
  \caption{IRAM-30m and PdBI maps of the Horsehead edge. Maps were
    rotated by $14^{\deg}$ counter-clockwise around the projection
    center, located at $(\delta x, \delta y) = (20'', 0'')$, to bring
    the exciting star direction in the horizontal direction. The
    horizontal zero, marked by the red vertical line, delineates the
    PDR edge. The crosses show the positions of the PDR (green) and
    the dense core (blue), where deep integrations were performed in
    the Horsehead WHISPER line survey (PI: J.Pety). The white lines
    delineate the arc-like structure of the \dcop{} emission. The
    spatial resolution is plotted in the bottom-left corner. Values of
    contour levels are shown in the respective image lookup table. The
    emission of all lines is integrated between 10.1 and 11.1 $\kms$}
  \label{fig:maps}
\end{figure*}
}

\newcommand{\FigLines}{%

\begin{figure*}[t!]
  \centering
  \includegraphics[width=\textwidth]{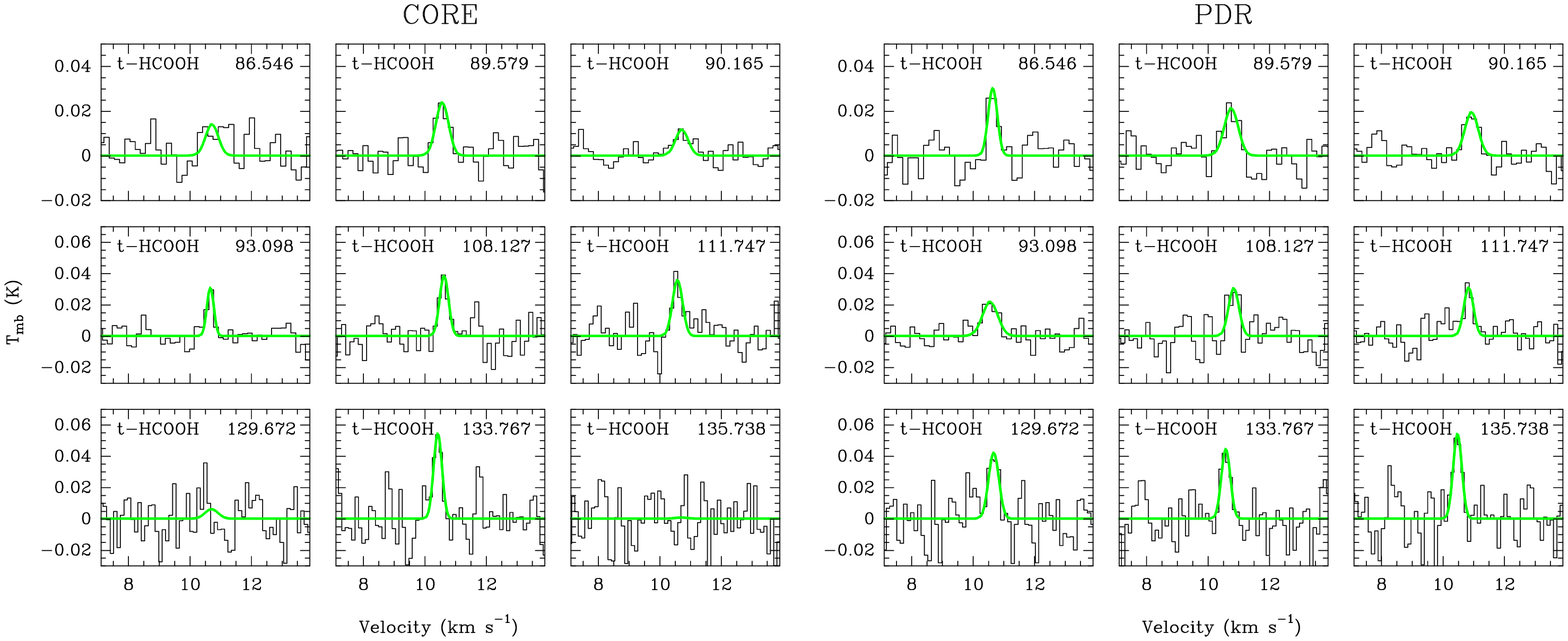}
  \caption[HCOOH lines detected toward the PDR and dense core]{HCOOH
    lines detected toward the dense core (\textit{left}) and PDR
    (\textit{right}). For each line, the same scale is used at both
    positions to ease the comparison.}
  \label{fig:hcooh-lines}
\end{figure*}

\begin{figure*}[t!]
  \centering
  \includegraphics[width=\textwidth]{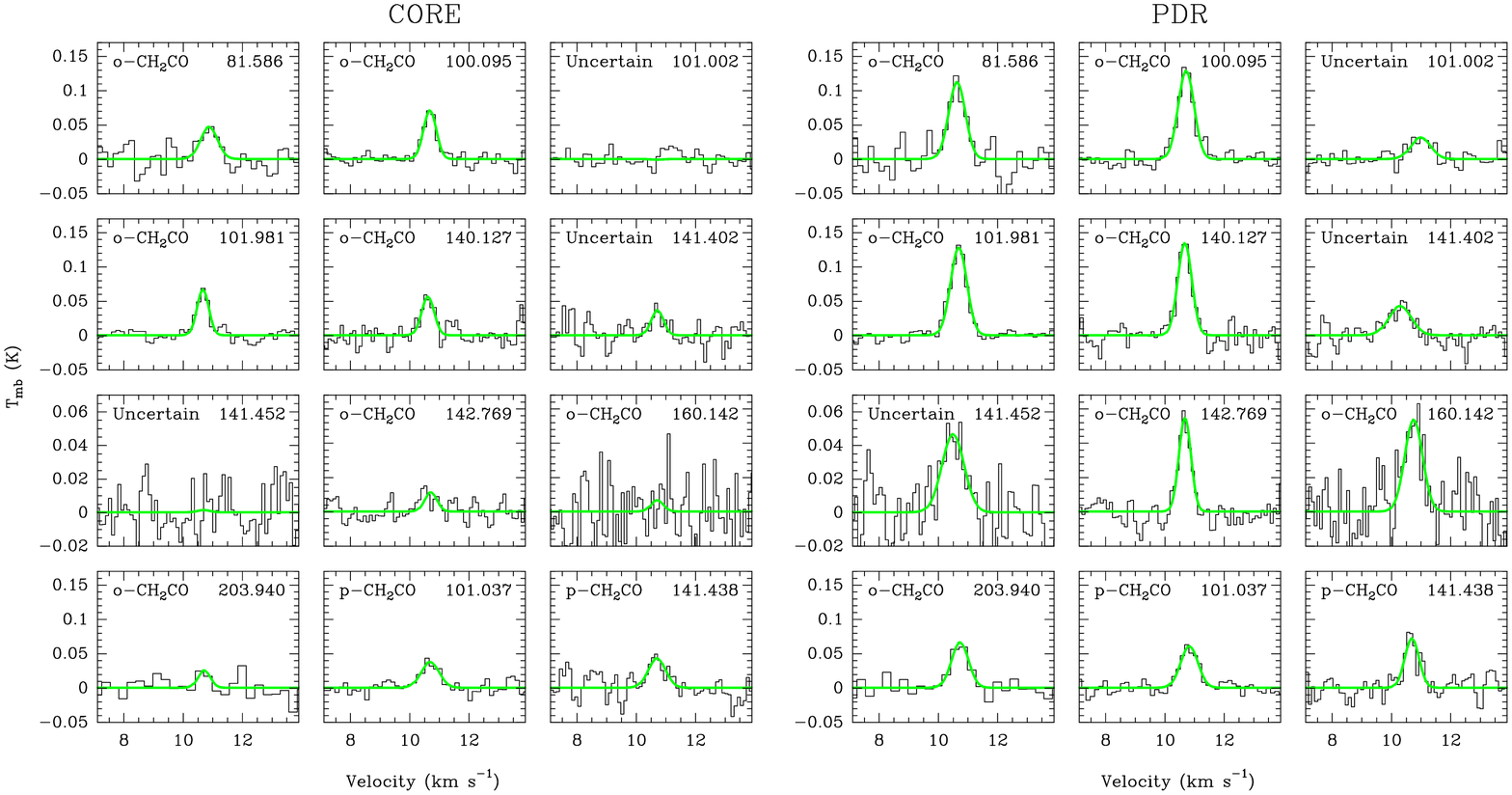} 
  \caption[\chhco{} lines detected toward the PDR and dense
    core]{\chhco{} lines detected toward the dense core (\textit{left}) and
    PDR (\textit{right}). For each line, the same scale is used
    at both positions to ease the comparison.}
  \label{fig:ch2co-lines}
\end{figure*}

\begin{figure*}[t!]
  \centering
  \includegraphics[width=\textwidth]{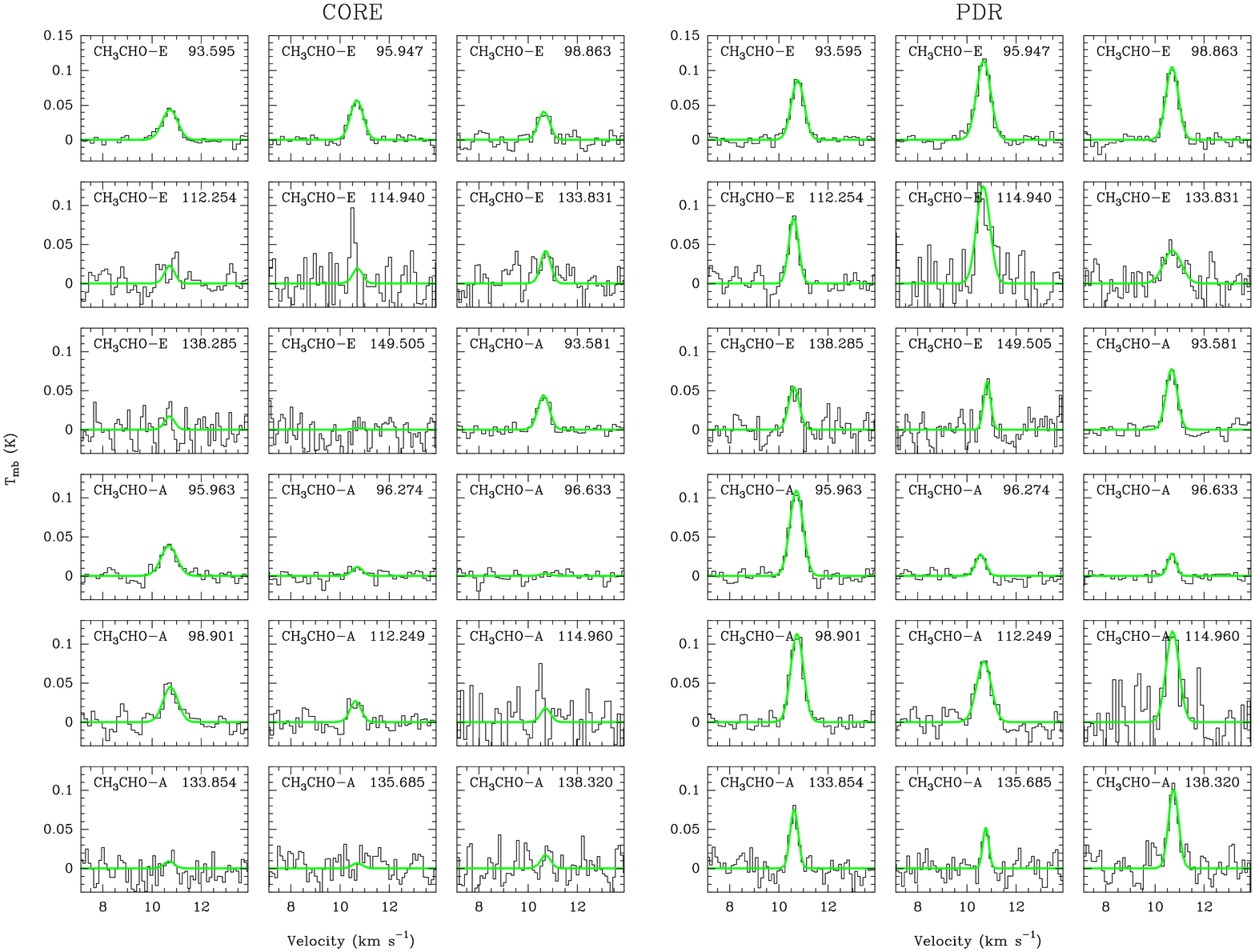} 
  \caption[\chhhcho{} lines detected toward the PDR and dense
    core]{\chhhcho{} lines detected toward the dense core (\textit{left})
    and PDR (\textit{right}). For each line, the same scale is
    used at both positions to ease the comparison.}
  \label{fig:ch3cho-lines}
\end{figure*}

\begin{figure*}[t!]
  \centering
  \includegraphics[width=\textwidth]{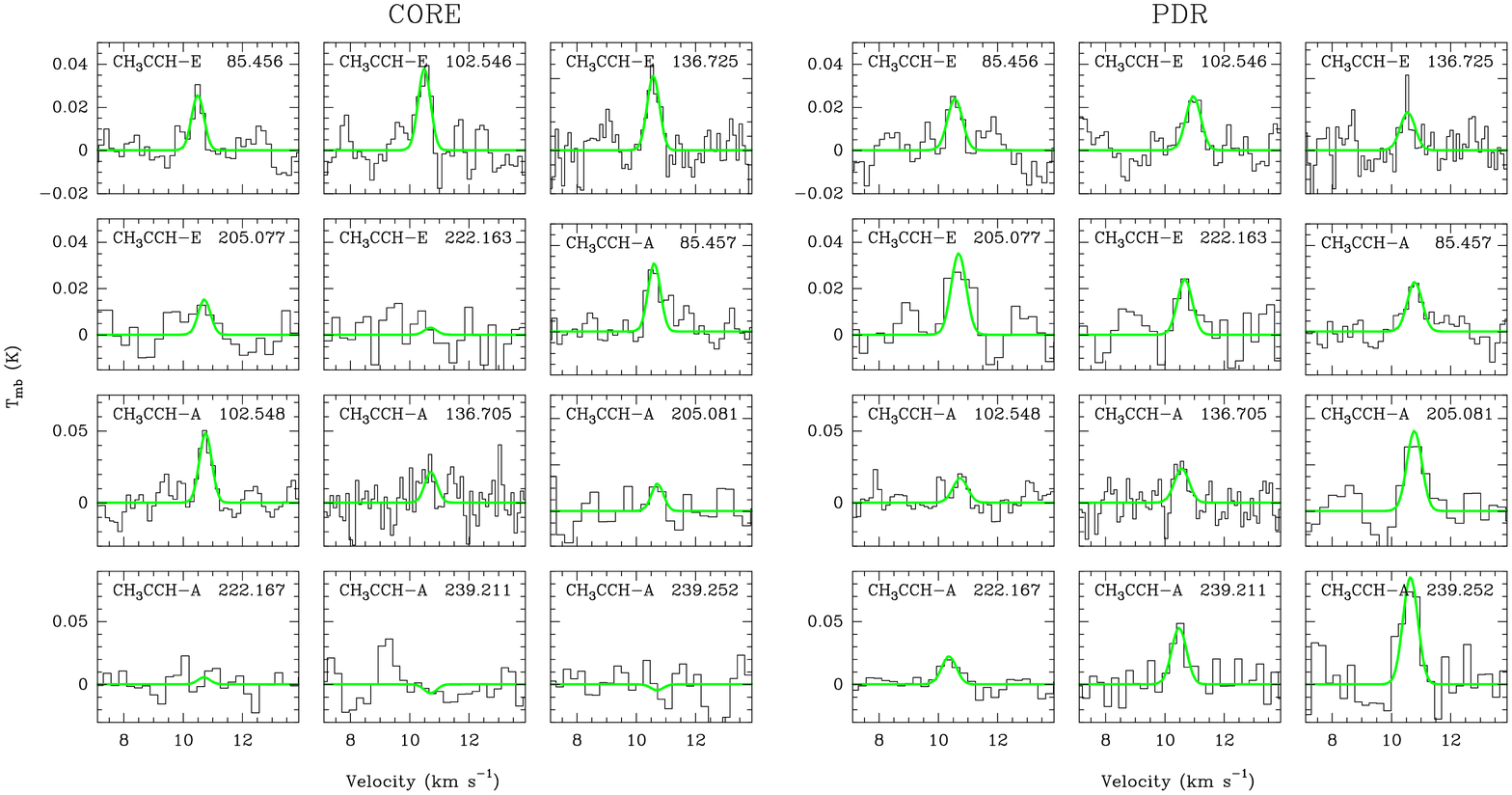} 
  \caption[\chhhcch{} lines detected toward the PDR and dense
    core]{\chhhcch{} lines detected toward the dense core (\textit{left})
    and PDR (\textit{right}). For each line, the same scale is
    used at both positions to ease the comparison.}
  \label{fig:ch3cch-lines}
\end{figure*}
}

\newcommand{\FigRotDiag}{%
\begin{figure*}[t!]
  \includegraphics[width=0.5\textwidth]{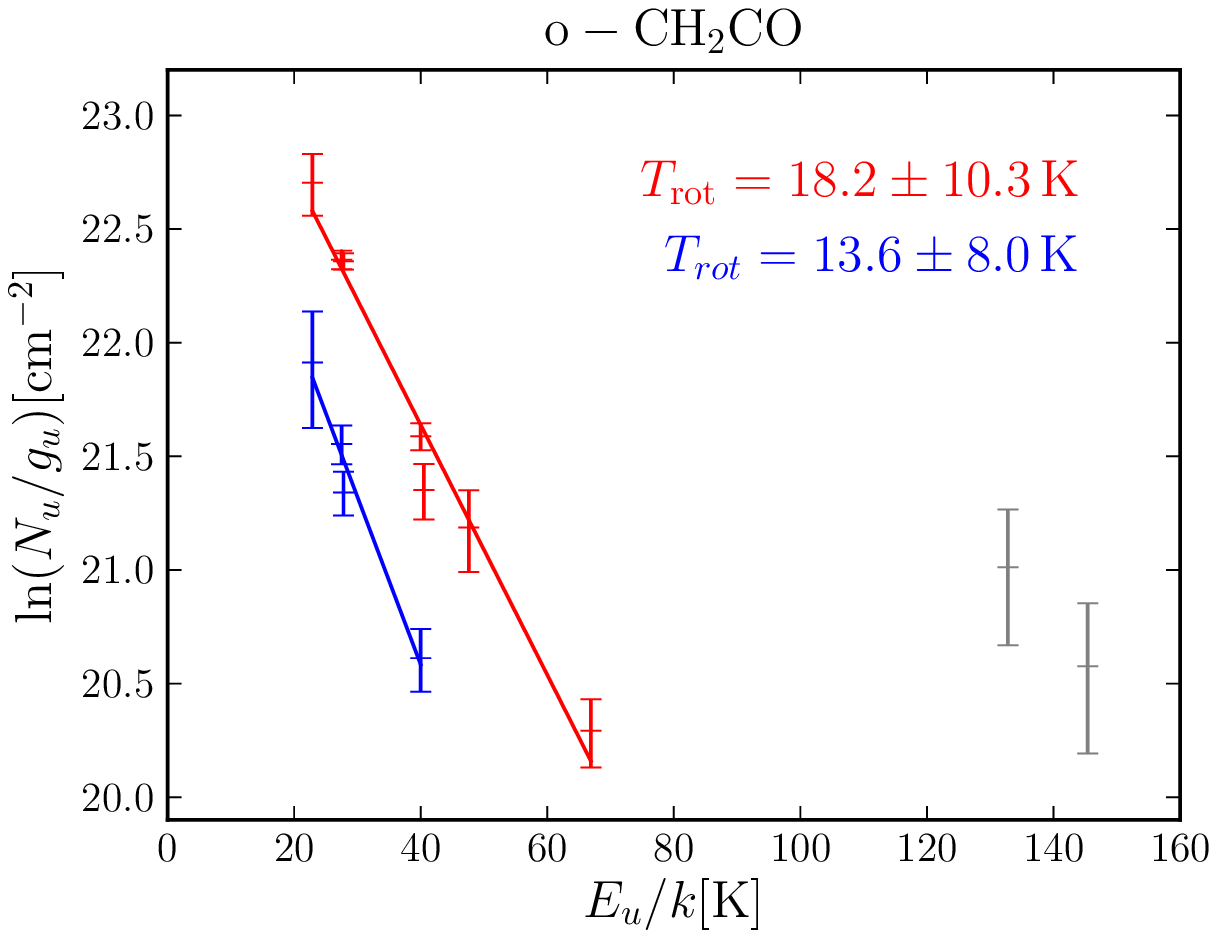}
  \includegraphics[width=0.5\textwidth]{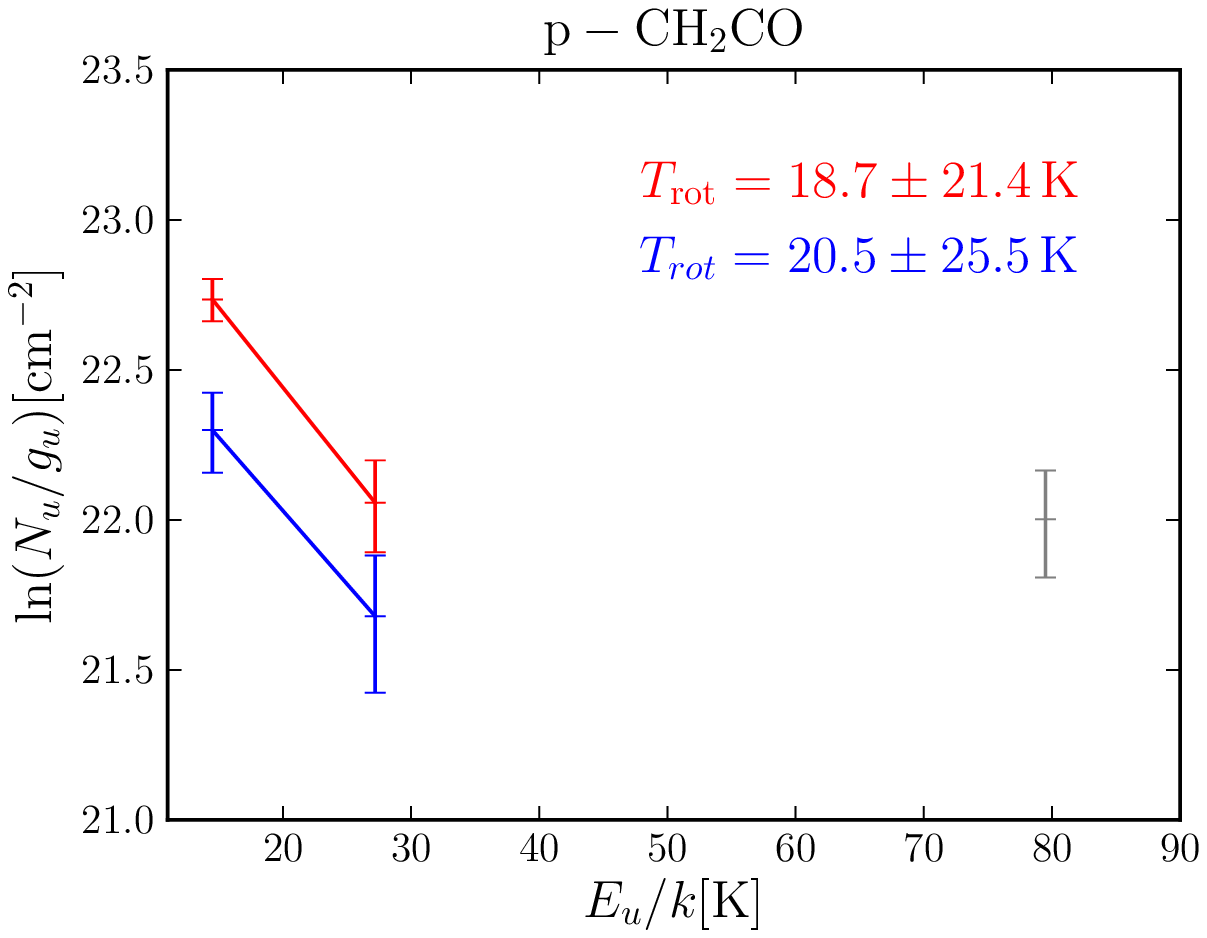}\\
  \includegraphics[width=0.5\textwidth]{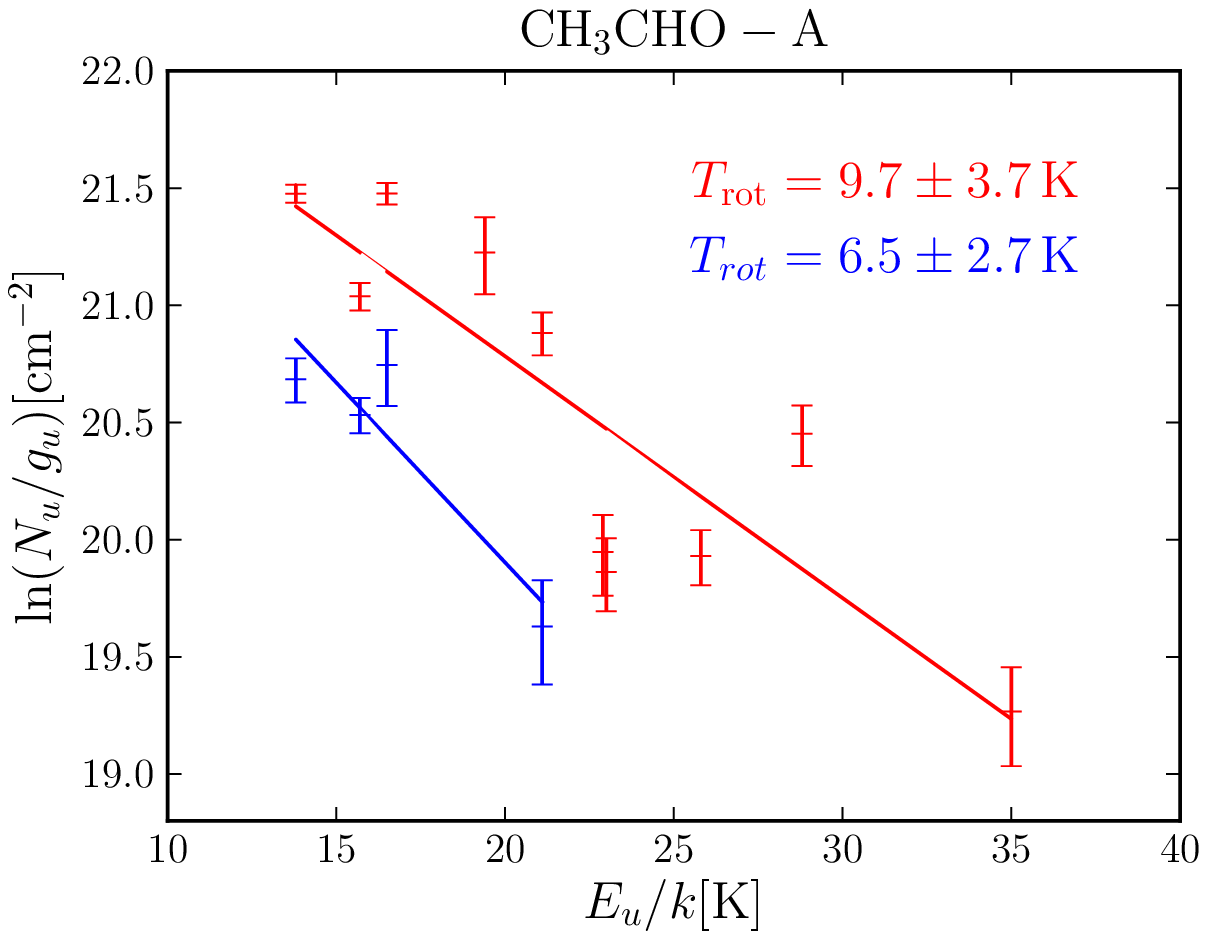}
  \includegraphics[width=0.5\textwidth]{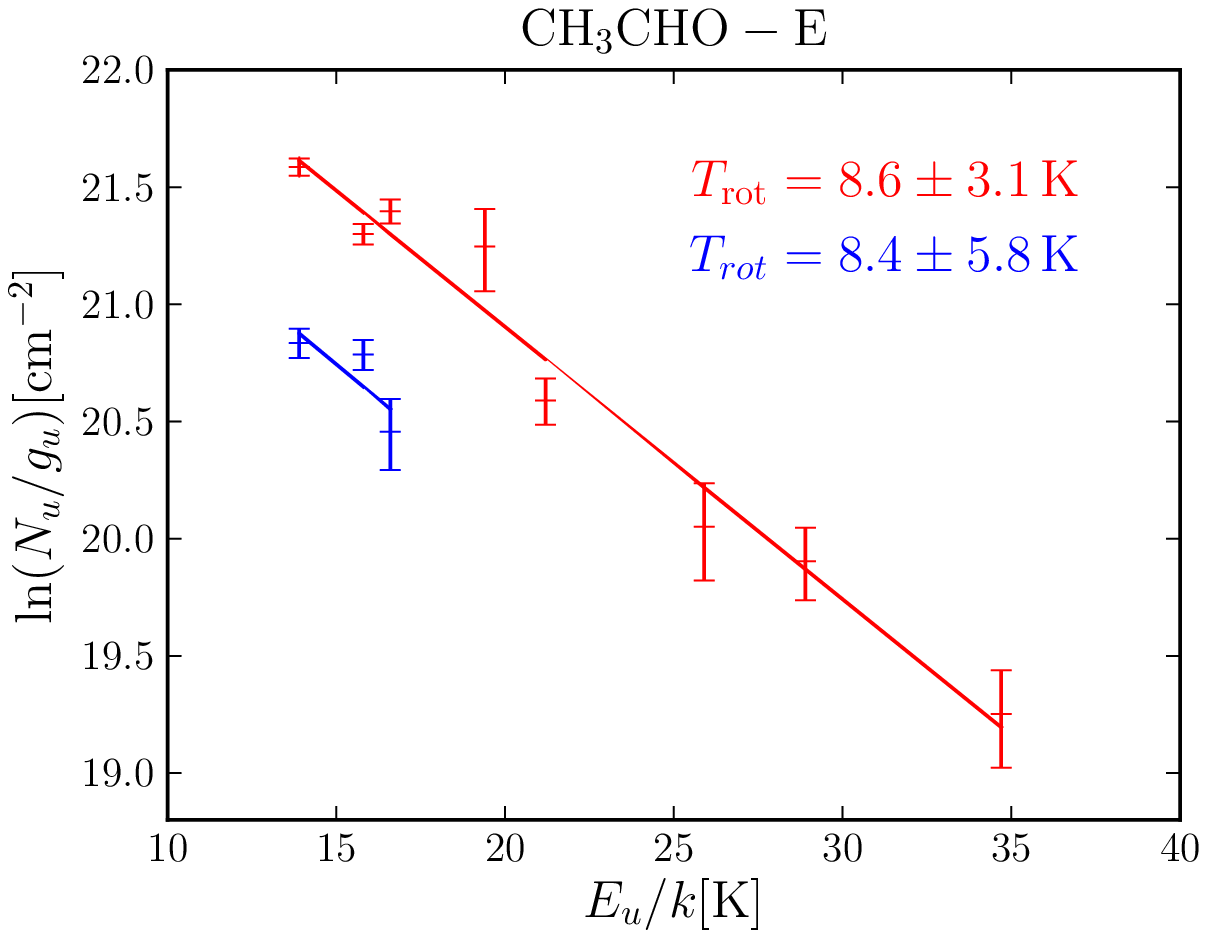}\\
  \includegraphics[width=0.5\textwidth]{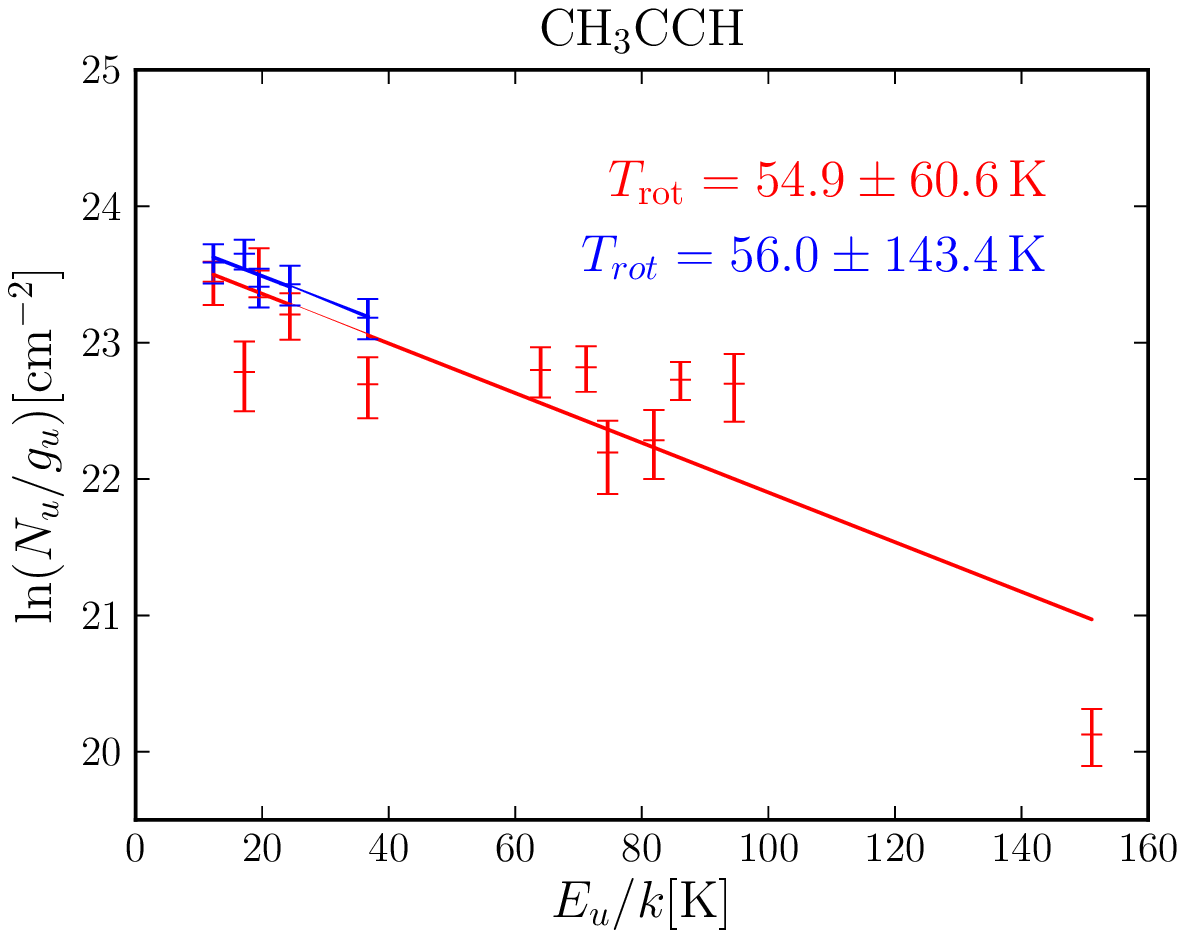}
  \includegraphics[width=0.5\textwidth]{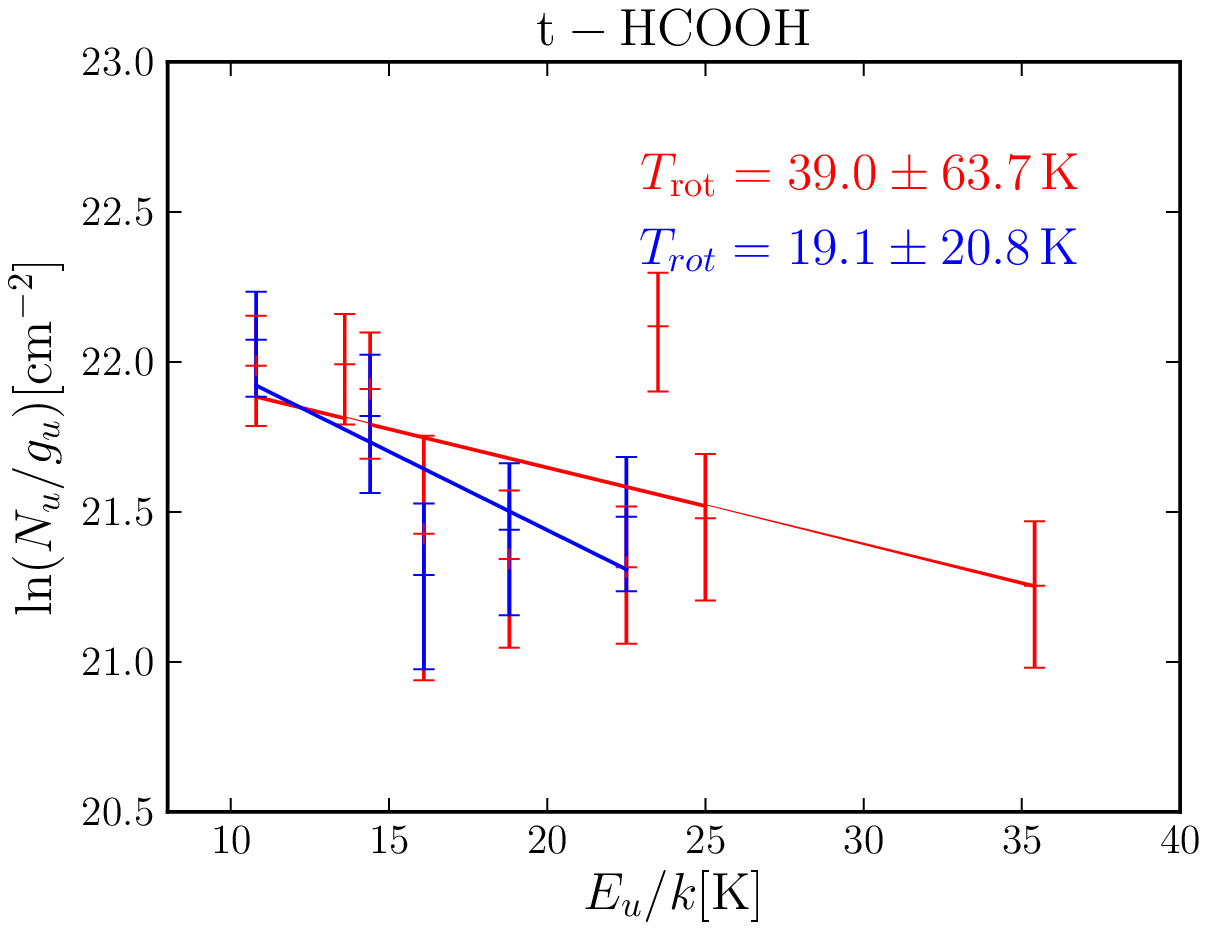}
  \caption[Rotational diagrams for HCOOH, \chhco{}, \chhhcho{} and
    \chhhcch{}]{Rotational diagrams for the PDR (red) and dense core
    (blue). The gray points correspond to the three lines detected at the
    PDR whose identification is uncertain. They are not considered in
    the fit.}
  \label{fig:rot-diag}
\end{figure*}
}

    %%%%%%%%%%%%%%%%%%%

\thispagestyle{plain}
\fancypagestyle{plain}{
\fancyhead[L]{\includegraphics[height=8pt]{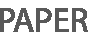}}
\fancyhead[C]{\hspace{-1cm}\includegraphics[height=15pt]{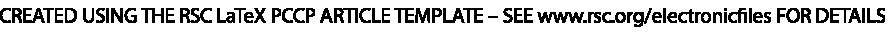}}
\fancyhead[R]{\includegraphics[height=10pt]{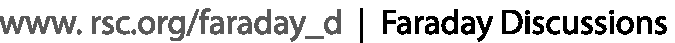}\vspace{-0.2cm}}
\renewcommand{\headrulewidth}{1pt}}
\renewcommand{\thefootnote}{\fnsymbol{footnote}}
\renewcommand\footnoterule{\vspace*{1pt}% 
\hrule width 11.3cm height 0.4pt \vspace*{5pt}} 
\setcounter{secnumdepth}{5}

\makeatletter 
\renewcommand{\fnum@figure}{\textbf{Fig.~\thefigure~~}}
\def\subsubsection{\@startsection{subsubsection}{3}{10pt}{-1.25ex plus -1ex minus -.1ex}{0ex plus 0ex}{\normalsize\bf}} 
\def\paragraph{\@startsection{paragraph}{4}{10pt}{-1.25ex plus -1ex minus -.1ex}{0ex plus 0ex}{\normalsize\textit}} 
\renewcommand\@biblabel[1]{#1}            
\renewcommand\@makefntext[1]% 
{\noindent\makebox[0pt][r]{\@thefnmark\,}#1}
\makeatother 
\sectionfont{\large}
\subsectionfont{\normalsize} 

\fancyfoot{}
\fancyfoot[LO,RE]{\vspace{-7pt}\includegraphics[height=8pt]{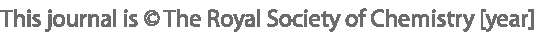}}
\fancyfoot[CO]{\vspace{-7pt}\hspace{5.9cm}\includegraphics[height=7pt]{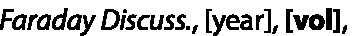}}
\fancyfoot[CE]{\vspace{-6.6pt}\hspace{-7.2cm}\includegraphics[height=7pt]{RF}}
\fancyfoot[RO]{\scriptsize{\sffamily{1--\pageref{LastPage} ~\textbar  \hspace{2pt}\thepage}}}
\fancyfoot[LE]{\scriptsize{\sffamily{\thepage~\textbar\hspace{3.3cm} 1--\pageref{LastPage}}}}
\fancyhead{}
\renewcommand{\headrulewidth}{1pt} 
\renewcommand{\footrulewidth}{1pt}
\setlength{\arrayrulewidth}{1pt}
\setlength{\columnsep}{6.5mm}
\setlength\bibsep{1pt}

\noindent\LARGE{\textbf{Chemical complexity in the Horsehead\\
    Photo-Dissociation Region}}
\vspace{0.6cm}

\noindent\large{\textbf{Viviana V. Guzm\'an \textit{$^{a}$}, J\'er\^ome
    Pety\textit {$^{a,b}$}, Pierre Gratier \textit{$^{a,b}$}, Javier
    R. Goicoechea \textit{$^{c}$}, Maryvonne Gerin \textit{$^{b}$},
    Evelyne Roueff \textit{$^{d}$}, Franck Le Petit \textit{$^{d}$} and Jacques Le
    Bourlot \textit{$^{d}$}}}\vspace{0.5cm}
%Please note that \ast indicates the corresponding author(s) but no footnote text is required. 

\noindent\textit{\small{\textbf{Received 29th November 2013, Accepted 23th January 2014}}}%\newline First published on the web Xth XXXXXXXXXX 200X}}}

\noindent \textbf{\small{DOI: 10.1039/c3fd00114h}}
\vspace{0.6cm}

\noindent \normalsize{The interstellar medium is known to be
  chemically complex. Organic molecules with up to 11 atoms have been
  detected in the interstellar medium, and are believed to be
  formed on the ices around dust grains. The ices can be released into
  the gas-phase either through thermal desorption, when a newly formed
  star heats the medium around it and completely evaporates the ices;
  or through non-thermal desorption mechanisms, such as
  photodesorption, when a single far-UV photon releases only a few
  molecules from the ices. The first one dominates in hot cores, hot
  corinos and strongly UV-illuminated PDRs, while the second one
  dominates in colder regions, such as low UV-field PDRs. This is the
  case of the Horsehead were dust temperatures are $\simeq20-30$~K,
  and therefore offers a clean environment to investigate what is the
  role of photodesorption. We have carried-out an unbiased spectral
  line survey at 3, 2 and 1mm with the IRAM-30m telescope in the
  Horsehead nebula, with an unprecedented combination of bandwidth,
  high spectral resolution and sensitivity. Two positions were
  observed: the warm PDR and a cold condensation shielded from the UV
  field (dense core), located just behind the PDR edge. We summarize
  our recently published results from this survey and present the
  first detection of the complex organic molecules \hcooh{}, \chhco{},
  \chhhcho{} and \chhhcch{} in a PDR. These species together with
  \chhhcn{} present enhanced abundances in the PDR compared to the
  dense core. This suggests that photodesorption is an efficient
  mechanism to release complex molecules into the gas-phase in far-UV
  illuminated regions.}

%The abstract should be a single paragraph which summarises the content of the article. Any references in the abstract should be written out in full \textit{e.g.} [Surname \textit{et al., Journal Title}, 2000, \textbf{35}, 3523].}
\vspace{0.5cm}

\section{Introduction}
%Footnotes
%\footnotetext{\dag~Electronic Supplementary Information (ESI) available: [details of any supplementary information available should be included here]. See DOI: 10.1039/c000000x/}

%Please use \dag to cite the ESI in the main text of the article.
%If you article does not have ESI please remove the the \dag symbol from the title and the above footnotetext.

\footnotetext{\textit{$^{a}$~IRAM, 300 rue de la Piscine, 38406 Saint
    Martin d'He\`res, France. E-mail: vguzman@cfa.harvard.edu}}
\footnotetext{\textit{$^{b}$~LERMA-LRA, UMR 8112, Observatoire de
    Paris and \'Ecole normale Sup\'erieure, 24 rue Lhomond, 75231 Paris,
    France}}
\footnotetext{\textit{$^{c}$~Centro de Astrobiolog\'ia, CSIC-INTA,
    Carretera de Ajalvir, Km 4, Torrej\'on de Ardoz, 28850 Madrid,
    Spain}} 
\footnotetext{\textit{$^{d}$~LUTH UMR 8102, CNRS and Observatoire de
    Paris, Place J. Janssen, 92195 Meudon Cedex, France}}

%Address, Address, Town, Country. Fax: XX XXXX XXXX; Tel: XX XXXX XXXX; E-mail: xxxx@aaa.bbb.ccc}}
%\footnotetext{\textit{$^{b}$~Address, Address, Town, Country. }}

%additional addresses can be cited as above using the lower-case letters, c, d, e... If all authors are from the same address, no letter is required

%\footnotetext{\ddag~Additional footnotes to the title and authors can be included \emph{e.g.}\ `Present address:' or `These authors contributed equally to this work' as above using the symbols: \ddag, \textsection, and \P. Please place the appropriate symbol next to the author's name and include a \texttt{\textbackslash footnotetext} entry in the the correct place in the list.}

Molecular lines are used to trace the structure of the interstellar
medium (ISM) and the physical conditions of the gas in different
environments, from high-z galaxies to proto-planetary disks. However,
the interpretation of molecular observations for most of these objects
is hampered by the complex source geometries, and the small angular
sizes in the sky compared with the angular resolution of current
instrumentation, that prevent us from resolving the different gas
components, and hence to know which specific region each molecule
actually traces. Therefore, in order to fully benefit from the
diagnostic power of the molecular lines, the formation and destruction
paths of the molecules must be quantitatively understood. This
challenging task requires the contribution of theoretical models,
laboratory experiments and observations. Well-defined sets of
observations of simple \textit{template} sources are key to benchmark
the predictions of theoretical models. In this respect, the Horsehead
nebula has proven to be a good template source of low-UV field
irradiated environments because it is close-by ($\sim400$~pc), it has
a simple geometry (edge-on) and its gas density is well
constrained. Moreover, in contrast to other Galactic
photo-dissociation regions (PDRs), like the Orion Bar and Mon R2 which
present large radiation fields ($\chi\simeq10^4-10^5$), the Horsehead
is illuminated by a weaker radiation field ($\chi\sim60$) and thus
better resembles the majority of the far-UV illuminated neutral gas in
the Galaxy. Furthermore, the dust grains in the Horsehead have
temperatures of $\simeq20-30$~K, which is not enough to thermally
desorb most of the ices. The Horsehead therefore offers a clean
environment to isolate the role of photo-desorption of ices on dust
grains.

Observations by the Infrared Space Observatory (ISO) and Spitzer have
shown that dust grains are covered by ice mantles in the cold
envelopes surrounding high-mass protostars\citep{gibb2000,gibb2004},
low-mass protostars
\citep{boogert2008,pontoppidan2008,Oberg2008,bottinelli2010} and in
isolated dense cores \citep{boogert2011}. These studies revealed that
the ice mantles consist mostly of \hho{}, CO$_2$ and CO, with smaller
amounts of \chhhoh{}, CH$_4$, NH$_3$ and \hhco{}. More complex
prebiotic molecules, such as glycine (NH$_2$CH$_2$COOH) could also
form on the ices around dust grains. Although their exact formation is
unclear, it is believed that the simplest prebiotic molecules have an
interstellar origin \citep{garrod2013,elsila2007}. Indeed, numerous
amino acids, which are the building blocks of proteins, have been
found in meteorites \citep{glavin2013}. In addition, glycine, which is
the simplest amino acid, has been detected in samples returned by
NASA's Stardust spacecraft from comet Wild~2
\citep{elsila2009}. Despite controversial detection claims
\citep{snyder2005,jones2007,cunningham2007}, glycine or other more
complex amino acids have not been detected in the interstellar medium
yet. The most complex molecules detected in the interstellar medium so
far, are glycolaldehyde\citep{hollis2000} (CH$_2$(OH)CHO),
acetamide\citep{hollis2006} (CH$_3$CONH$_2$),
aminoacetonitrile\citep{belloche2008} (NH$_2$CH$_2$CN), and the ethyl
formate\citep{belloche2009} (C$_2$H$_5$OCHO). This shows the high
degree of chemical complexity that can be reached in the interstellar
medium.

\TabObsMaps{}

Simpler, but still complex organic molecules, such as methanol
(\chhhoh{}), ketene (\chhco{}), acetaldehyde (\chhhcho{}), formic acid
(HCOOH), formamide (NH$_2$CHO), propyne (\chhhcch{}), methyl formate
(HCOOCH$_3$), and dimethyl ether (CH$_3$OCH$_3$), are widely observed in
hot cores of high-mass protostars
\citep{cummins1986,blake1987,bisschop2007}, and also in hot-corinos of
low-mass protostars \citep{vandishoeck1995,cazaux2003}. The complex
molecules observed in protostars have been classified in three
different generations by \citet{herbst2009}, depending on their
formation mechanism. The zeroth generation species form through grain
surface processes in the cold ($<20$~K) pre-stellar stage (\eg{},
\hhco{} and \chhhoh). First generation species form from surface
reactions between photodissociated products of the zeroth generation
species in the warm-up ($20-100$~K) period. Finally, second
generations species form in the hot ($>100$~K) gas from the evaporated
zeroth and first generation species in the so called hot-core
phase. Although it is clear that grain surface processes play an
important role in the formation of complex molecules, the exact
formation mechanism of most complex molecules is still debated.

\TabAbundances{}
\FigMaps{}

\citet{bisschop2007} observed several complex molecules toward seven
high-mass protostars, and classified them as cold (T$<100$~K) and hot
(T$>100$~K) molecules based on their rotational temperatures. The hot
molecules include \hhco{}, \chhhoh{}, HNCO, \chhhcn{}, HCOOCH$_3$ and
CH$_3$OCH$_3$, while the cold molecules include HCOOH, \chhco{},
\chhhcho, and \chhhcch. The cold molecules are expected to be present
in the colder envelope around the hot-core. \citet{oberg2013} studied
the spatial distribution of complex molecules around a high-mass
protostar and found that \chhco{}, \chhhcho{} and \chhhcch{} are
indeed abundant in the cold envelope. They classified them as zeroth
order molecules because their formation must require very little heat.

Complex organic molecules may trace other environments than hot cores
and hot corinos. They are also present in the cold UV-shielded
gas. \citet{bacmann2012} detected CH$_3$OCH$_3$, CH$_3$OCHO,
\chhco{} and \chhhcho{} in a cold ($\Tkin\sim10$~K) prestellar core. These
observations challenged the current formation scenario of complex
molecules on dust grains, because the diffusion reactions that lead to
the formation of species are not efficient on dust grains with
temperatures of $\sim10$~K. \chhhcho{} and \chhco{} have also been
detected in the dark cloud TMC-1
\citep{matthews1985,irvine1989}. \chhco{} and \chhhcho{} have also
been detected in a $z=0.89$ spiral galaxy located in front of the
quasar PKS1830-211 \citep{muller2011}.

In this paper, we present the results of an unbiased line survey
performed with the IRAM-30m telescope in a classic star forming
region, the Horsehead nebula. We describe the observations in
section~\ref{sec:obs}. In section~\ref{sec:results} we present a
summary of the recently published results of the line survey. In
section~\ref{sec:complex} we present new unpublished results about the
first detection of complex molecules in a PDR. We discuss these
observations in section~\ref{sec:discussion} and conclude in
section~\ref{sec:conclusions}.

%Complex molecules have been detected in several environments including
%cold dense cores and protostellar sources. Here, we present the first
%detection of complex molecules in a PDR.

\section{Observations}
\label{sec:obs}

\subsection{Deep pointed integrations: The Horsehead WHISPER line survey}

With the purpose of providing a benchmark to chemical models we have
performed a complete and unbiased line survey: the Horsehead WHISPER
(Wideband High-resolution Iram-30m Surveys at two Positions with Emir
Receivers, PI: J. Pety). Two positions were observed: 1) the HCO peak,
which is characteristic of the photo-dissociation region at the
surface of the Horsehead nebula \citep{gerin2009}, and 2) the $\dcop$
peak, which belongs to a cold condensation located less than $40''$
away from the PDR edge, where HCO$^+$ and other species are highly
deuterated \citep{pety2007}. Hereafter we refer to these two positions
as the PDR and dense core, respectively. The combination of the new
EMIR receivers at the IRAM-30m telescope and the Fourier transform
spectrometers (FTS) yields a spectral survey with unprecedented
combination of bandwidth (36~GHz at 3mm, 34~GHz at 2mm and 73~GHz at
1mm), spectral resolution (49~kHz at 3 and 2mm; and 195~kHz at 1mm),
and sensitivity (median noise 8.1~mK, 18.5~mK and 8.6~mK at 3, 2 ans
1mm respectively). A detailed presentation of the observing strategy
and data reduction process will be given in a forthcoming paper. In
short, all frequencies were observed with two different frequency
tunings and the Horsehead PDR and dense core positions were
alternatively observed every 15 minutes in position-switching mode
with a common fixed off-position. This observing strategy allows us to
remove potential ghost lines that are incompletely rejected from a
strong line in the image sideband (the typical rejection of the EMIR
sideband-separating mixers is only 13dB).
 
The line density at 3~mm is, on average, 5 and 4~lines/GHz in the PDR
and dense core, respectively. At 2 and 1~mm, the line density is
1~line/GHz in both the PDR and dense core. The contribution of
molecular lines to the total flux at 1.2 mm is estimated to be 14\% at
the PDR and 16\% at the dense core. Approximately 30 species (plus
their isotopologues) are detected with up to 7 atoms in the PDR and
the dense core. 

\subsection{IRAM-30m and PdBI maps}

Figure~\ref{fig:maps} displays the integrated emission of the \cch{},
\chhhcho{}, HCO, \cfp{}, \dcop{}, \phhco{}, and \chhhohA{} lines as
well as the 1.2mm continuum emission. The observation parameters are
summarized in Table~\ref{tab:obs:maps}. A reference is given where a
detailed description of the observations and data reduction can be
found for each map.
 
The A- and E-type \chhhcho{} $5_{15}-4_{14}$ lines at 93.581~GHz and
93.595~GHz were observed simultaneously with the A- and E-type
\chhhcho{} $6_{16}-5_{15}$ lines at 112.249~GHz and 112.254~GHz during
$\sim 17$ hours of average summer weather in August and
September 2013. We used the two polarizations of the EMIR receivers
and the FTS backends at 49~kHz spectral resolution. We used the
position-switching, on-the-fly observing mode. The off-position
offsets were ($\delta$RA$,\delta$Dec) = $(100'', 0'')$, that is, the
{\sc{H\,ii}} region ionized by $\sigma$Ori and free of molecular
emission. We observed along and perpendicular to the direction of the
exciting star in zigzags (\ie{}, $\pm$ the lambda and beta scanning
direction). From our knowledge of the IRAM-30m telescope, we estimate
the absolute position accuracy to be $3''$.

The IRAM-30m data were processed with the \GILDAS{}/\CLASS{}
software. The data were first calibrated to the \Tas{} scale using the
chopper-wheel method \citep{penzias1973}. The data were converted to
main-beam temperatures (\Tmb{}) using the forward and main-beam
efficiencies (\Feff{} and \Beff{}). The resulting amplitude accuracy
is 10\%. We then computed the experimental noise by subtracting a
zeroth-order baseline from every spectra. A systematic comparison of
this noise value with the theoretical noise computed from the system
temperature, the integration time, and the channel width allowed us to
filter out outlier spectra. The spectra where then gridded to a data
cube through a convolution with a Gaussian kernel. In order to
increase the signal-to-noise ratio, we smoothed the four A- and E-type
\chhhcho{} lines to the largest angular resolution and then averaged
all the data. The averaged map, which has a final resolution of
$30''$, is shown in Fig~\ref{fig:maps}.

\section{Recent results from the Horsehead WHISPER line survey}
\label{sec:results}

\subsection{CF$^\mathrm{+}$: a tracer of C$^\mathrm{+}$ and a
  measure of the fluorine abundance}

\cfp{}, which was only detected in the Orion Bar before
\citep{neufeld2006}, was detected toward the illuminated edge of the
Horsehead nebula by \citet{guzman2012a}. The \cfp{} ion, which is
formed by reactions of HF and \cp{}, is expected to be the second most
important fluorine reservoir, after HF, in regions where \cp{} is
abundant \citep{neufeld2005}. Indeed, the \cfp{} emission is
concentrated toward the edge of the Horsehead, delineating the western
edge of the \dcop{} emission (see Fig.~\ref{fig:maps}). Theoretical
models predict that there is a significant overlap between \cfp{} and
\cp{} at the edges of molecular clouds. Therefore, we propose that
\cfp{} can be used as a proxy of \cp{}, but that can be observed from
ground-based telescopes, unlike \cp{} for which we need to go to
space. This can be a powerful tool, because the [C\,{\sc ii}]
157.8~$\mu$m line is the main cooling mechanism of the diffuse gas,
and the cooling of the medium, which allows the gas to compress, is a
crucial step in the formation of new stars. Moreover, given the simple
chemistry of fluorine and assuming that the \cfp{} destruction is
dominated by dissociative recombination with electrons with little
contribution from photodissociation (which is true only in low-UV
PDRs), one obtains that the \cfp{} column density is proportional to
the column density of HF. Then, assuming that in molecular clouds all
fluorine is in its molecular form, the elemental abundance of fluorine
can be derived directly from \cfp{} observations. We infer
F/H$=(0.6-1.5)\times10^{-8}$ in good agreement with the one found in
diffuse molecular clouds \citep{sonnentrucker2010}, and somewhat lower
than the solar value \citep{asplund2009} and the one found in the
diffuse atomic gas \citep{snow2007}. Finally, because the Horsehead
shows narrow emission lines, in contrast to other PDRs like the
Orion~Bar, \citet{guzman2012b} were able to resolve the two hyperfine
components in the \cfp{} $J=1-0$ line and to compare with \emph{ab
  initio} computations of the \cfp{} spin rotation constant. The
derived theoretical value of $C_I = 229.2$~kHz agrees well with the
observations. The Horsehead is thus a good laboratory for precise
spectroscopic studies of species present in far-UV illuminated
environments.

\subsection{Detection of a new molecule in space, tentatively
  attributed to l-C$_\mathbf{3}$H$^\mathbf{+}$}

Thanks to the sensitive observations and large bandwidth covered by
the Horsehead WHISPER line survey, a consistent set of 8 lines were
detected toward the PDR position, that could not be associated
to any transition listed in the public line catalogs. The observed
lines can be well fitted with a linear rotor model, implying a
closed-shell molecule. The deduced rotational constant value is close
to that of \ccch{}. In addition, the spatial distribution of the
species integrated emission has a shape similar to radical species
such as HCO, and small hydrocarbons such as \cch{} (see
Fig.~\ref{fig:maps}). Therefore, \citet{pety2012} attributed the
detected lines to the small hydrocarbon cation l-\ccchp{}.

In the family of small hydrocarbons, \citet{pety2005} found that
\ccch{} and \ccchh{} are about 1 order of magnitude more abundant in
PDRs than current pure gas-phase models predict. An additional
formation mechanism is therefore needed. One possibility to explain
the observed high abundance of hydrocarbons in PDRs is the so called
\textit{Top-Down} model. In this scenario polycyclic aromatic
hydrocarbons (PAHs) are fragmented into small hydrocarbons in PDRs due
to the strong UV fields \citep{fuente2003,teyssier2004,pety2005}. In
the same way, PAHs are formed by photo-evaporation of very small
grains \citep{berne2007,pilleri2012}. The discovery of C$_3$H$^+$,
which is an intermediate species in the gas-phase formation scenario,
brings further constraints to the formation pathways of the small
hydrocarbons. Indeed, we find a l-C$_3$H$^+$ abundance which is too
low to explain the observed abundance of the other related small
hydrocarbons by means of pure gas-phase chemical reactions.

The lines detected in the Horsehead have been detected in other
environments, like the Orion Bar (Cuadrado et al. in prep) and Sgr~B2
\citep{mcguire2013}, which confirms the presence of the carrier in the
ISM. But the attribution of the unidentified lines to l-\ccchp{} has
been questioned by \citet{huang2013}, because their theoretical
calculations of the spectroscopic constants of \ccchp{} differ from
the ones inferred from our observations. \citet{fortenberry2013}
proposed that a more plausible candidate is the hydrocarbon anion
$\ccch^-$. However, if the unknown species is the anion, it would be
the first anion detected in the Horsehead, and the ratio of $\ccch^-$
to neutral \ccch{} would be $\sim$57\%, which is higher than any anion
to neutral ratio detected in the ISM so far.  In addition, the lines
were not detected in the dark cloud TMC~1, where other anions have
been already detected. Moreover, because $\ccch^-$ is an asymmetric
rotor, the lines detected in the Horsehead would correspond to the
$K_a=0$ ladder and the lines from the $K_a=1$ ladder should also be
detected. We find no evidence of the $K_a=1$ lines of $\ccch^−$ in the
observations of the Horsehead PDR \citep{mcguire2014}. For all these
reasons it would be unexpected that the carrier of the unidentified
lines is the anion, $\ccch^-$. The observations favor the assignment
of the unidentified species to the hydrocarbon cation, \ccchp{}, as
the most likely candidate. However, A direct measurement in the
laboratory is necessary to provide a definitive answer and close the
controversy created by these observations in the Horsehead. Ongoing
high-angular PdBI observations of this species in the Horsehead PDR
will allow us to better constrain the chemistry of small hydrocarbons
in the near future.

\subsection{Photo-desorption of dust grain ice mantles: H$_2$CO and CH$_3$OH}

\FigLines{}

Relatively simple organic molecules, like \hhco{} and \chhhoh{}, are
key species in the synthesis of more complex molecules in the ISM
\citep{bernstein2002,munoz-caro2002,garrod2008}, that could eventually
end up in proto-planetary disks, and hence in new planetary
systems. They are also used to probe the temperature and density of
the gas in different astrophysical
sources \citep{mangum1993,leurini2004,mangum2013}. Both \hhco{} and
\chhhoh{} have been detected in a wide range of interstellar
environments such as dark clouds, proto-stellar cores and comets, with
high abundances ($10^{-6}-10^{-9}$) with respect to total
hydrogen. Unlike \hhco{}, which can be formed efficiently in both the
gas-phase and on the surfaces of dust grains, \chhhoh{} is thought to
be formed mostly on the ices, through the successive additions of
hydrogen atoms to adsorbed CO molecules.

\citet{guzman2011} and \citet{guzman2013} observed several millimeter
lines of \hhco{} and \chhhoh{} toward the PDR and dense core positions
in the Horsehead. The inferred abundances from the observations (see
Table~\ref{tab:abundances}) were compared to PDR models that include
either pure gas-phase chemistry or both gas-phase and grain surface
chemistry. Pure gas-phase models cannot reproduce the observed
abundances of either \hhco{} or \chhhoh{} at the PDR position. Both
species are therefore mostly formed on the surface of dust grains,
probably through the successive hydrogenation of CO ices and are
subsequently released into the gas-phase through photodesorption. At
the dense core, on the other hand, photodesorption of ices is needed
to explain the observed abundance of \chhhoh, while a pure gas-phase
model can reproduce the observed \hhco{} abundance. The different
formation routes for \hhco{} at the PDR and dense core suggested by
the models is strengthened by the different ortho-to-para ratios
derived from the observations ($\sim3$ at the dense core, $\sim$2 at
the PDR).
 
In addition to the lines detected in the WHISPER survey, we obtained
high-angular resolution ($6''$) maps of \hhco{} and \chhhoh{} with the
IRAM-PdBI. Fig.~\ref{fig:maps} shows the \hhco{} and \chhhoh{}
single-dish 30m maps that were used for the short-spacing of the PdBI
observations. The \hhco{} emission map presents a peak at the dense
core position, while \chhhoh{} presents a dip in its emission at the
same position. The observations thus suggest that \chhhoh{} is
depleted in the dense core. This way, gas-phase \chhhoh{} is present
in an envelope around the dense core, while \hhco{} is present in both
the envelope and the dense core itself. Indeed, we expect
photodesorption to be more efficient at the PDR than at the
far-UV shielded dense core. We thus conclude that photo-desorption of
ices is an efficient mechanism to release species into the gas-phase
in far-UV illuminated regions.

\subsection{Nitrile molecules: CH$_\mathbf{3}$CN, CH$_\mathbf{3}$NC
  and HC$_\mathbf{3}$N} 

Moderately complex nitriles like \chhhcn{} and \hcccn{} are easily
detected in (massive) star forming regions. In particular, the
\chhhcn{} emission has been found to be enhanced in star forming
regions containing an ultracompact {\sc{H\,ii}}
region\citep{purcell2006}.  \chhhcn{} is thought to be a good tracer
of the physical conditions in warm and dense
regions. \citet{gratier2013} detected several lines of \chhhcn{} and
\hcccn{} in the PDR and dense core. \chhhnc{} and C$_3$N are also
detected toward the PDR. The observations show that the chemistry of
\hcccn{} and \chhhcn{} is quite different. Indeed, we find that
\chhhcn{} is 30 times more abundant in the far-UV illuminated gas than
in the far-UV shielded core, while \hcccn{} has a similar abundance in
both positions. The high abundance of \chhhcn{} inferred in the PDR is
surprising because the photodissociation of this complex molecule is
expected to be efficient in far-UV illuminated regions. The observed
abundance in the PDR cannot be reproduced by current pure gas-phase
chemical models. We have shown that photodesorption is an efficient
mechanism to release \hhco{} and \chhhoh{} in the PDR, but the case of
\chhhcn{} is even more extreme as it is 30 times more abundant there
than in the dense core, while \hhco{} presents similar abundances in
both positions. This shows that there is something specific in the
chemistry of \chhhcn{} in FUV-illuminated regions. \chhhcn{} could be
produced on the ices by the photo-processing of N bearing species
followed by photodesorption, but it could also be produced in the
gas-phase if the abundances of its gas-phase precursors, HCN and
CH$^+_3$, are enhanced. The detection of \chhhnc{} at the
PDR, which results in an \chhhnc{}/\chhhcn{} isomeric ratio of 0.15,
suggests that \chhhnc{} could also form on the surfaces of dust grains
through far-UV irradiation of \chhhcn{} ices leading to isomerization.

\section{Other complex molecules in PDRs}
\label{sec:complex}

Within the WHISPER line survey, several lines of HCOOH, \chhco{},
\chhhcho{} and \chhhcch{} are detected. These are presented in
Figs.~\ref{fig:hcooh-lines} to \ref{fig:ch2co-lines}. The
spectroscopic parameters and Gaussian fit results of the detected
lines are listed in Appendix~\ref{app:tables}. The spectroscopic
parameters are taken from the CDMS \citep{muller2001} and JPL
\citep{Pickett1998} data bases. Several lines of ketene and
acetaldehyde are clearly detected ($S/N > 5\sigma$). The formic acid
and propyne present several but fainter ($2\sigma-5\sigma$) lines
toward the PDR and dense core positions. In order to confirm the
correct identification of these molecules, we have modeled the
spectrum of each species assuming LTE and optically thin emission, and
checked that there are no predicted lines missing in the line
survey. Both ortho and para forms of \chhco{} are detected, as well as
both E and A forms of \chhhcho{} and \chhhcch{}. All the lines
detected of the formic acid correspond to the trans isomer. \chhco{},
\chhhcho{} lines are brighter toward the PDR position than toward the
dense core, while HCOOH and \chhhcch{} lines have similar brightness
in both positions.

The beam-averaged column density of each molecule was estimated using
rotational diagrams because no collisional coefficients are
available. The detected lines cover a sufficiently large energy range
to derive a rotational temperature. The resulting rotational diagrams
are shown in Fig.~\ref{fig:rot-diag}. The inferred abundances with
respect to H nuclei are summarized in Table~\ref{tab:abundances}. The
partition function was computed independently for ortho and para
nuclear spin forms (for ketene), and for E and A symmetry forms (for
\chhhcho{}), by direct summation over the energy levels.

\subsection*{Formic acid}
The data in the rotational diagram of HCOOH present a large scatter
because all the detected lines are weak and therefore have a larger
uncertainty than the lines of the other complex molecules. The
rotational temperature is poorly constrained. The inferred abundances
of $\sim5\times10^{-11}$ (PDR) and $\sim1\times10^{-11}$ (core) are
thus also uncertain and should be considered as an order of magnitude
estimate. Deeper integration times are needed to better constrain the
HCOOH abundance.

\subsection*{Ketene} 
The ortho and para symmetries of \chhco{} were treated as different
species. When including all the \ochhco{} and \pchhco{} lines detected
in the PDR, the fit results in rotational temperatures of 154~K and
120~K for \ochhco{} and \pchhco{}, respectively. These temperatures are
much larger than the kinetic temperature at the PDR ($\sim$60~K). When
the three lines of \chhco{} with energies above 80~K that are detected
at the PDR (gray points in Fig.~\ref{fig:rot-diag}) are not considered
in the rotational diagrams, the derived rotational temperatures
decrease to 18~K for both ortho and para species. This temperature
agrees much better with the expected sub-thermal excitation in the
Horsehead, and also with the derived rotational temperatures at the
dense core. The enhanced emission of the three lines with $E_u>80$~K
could be the result of an excitation effect. However, these lines are
broader (0.8~\kms) than the other \chhco{} lines and than other
species detected in the Horsehead PDR (the typical linewidth is
0.6~\kms{}). In addition, the velocity of these three lines differs by
0.2~\kms{} from the systemic velocity of 10.7~\kms{} found for most
species in the Horsehead.
%For these reasons, the identification of these three lines is
%uncertain. 
\citet{cummins1986} detected several \ochhco{} lines toward Sgr~B2,
including the $5_{33}-4_{32}$ at 101.002~GHz, which is one of the
three lines detected in the Horsehead with $E_u>80~K$. They also
obtained a large rotational temperature, which led them to remove this
line from the fit and consider the identification as uncertain. At the
dense core position in the Horsehead, the derived rotational temperature
for \ochhco{} is $\sim14$~K. Only two lines of \pchhco{} are detected
at the dense core, resulting in a rotational temperature of
$\sim20$~K. Ketene is $\sim3$ times more abundant in the PDR than in
the dense core, with abundances of $1.5\times10^{-10}$ (PDR) and
$4.9\times10^{-11}$ (core). The ortho-to-para ratio is poorly
constrained, resulting in $o/p=7.1\pm4.2$ (PDR) and $o/p=5.7\pm3.9$
(dense core).

\TabDipole{}

\FigRotDiag{}

\subsection*{Acetaldehyde}
The E and A symmetries of \chhhcho{} were also treated as different
species. Acetaldehyde has a large dipole moment (2.7~Debye, see
Table~\ref{tab:dipole}) compared to methanol (1.7~Debye). Since the
critical density is proportional to $\mu^2$, sub-thermal effects are
important for \chhhcho{} \citep{bisschop2007}. Indeed, the derived
rotational temperatures for \chhhcho{} ($6-10$~K) are the lowest ones
of all the molecules discussed here. The derived column density of
\chhhcho{} is $\sim3$ times larger in the PDR than in the dense
core. The inferred E/A ratio is $\sim0.3\pm0.1$ (PDR) and
$\sim0.2\pm0.2$ (core), \ie{}, much lower than unity in both
positions.

Figure~\ref{fig:maps} displays the averaged E and A \chhhcho{} lines
at 93.6~GHz. The \chhhcho{} emission clearly peaks at the PDR
position, delineating the edge of the Horsehead nebula. The \chhhcho{}
emission resembles the HCO emission at the 30m telescope angular
resolution of $30''$, which is concentrated in a narrow structure
peaking at the PDR. Higher-angular resolution ($6''$) observations by
\citet{gerin2009} showed that the HCO emission traces a filament of
$\sim12''$ width. The similarities between the emission of HCO and
\chhhcho{} thus suggest that \chhhcho{} also arises from a narrow
filament that peaks at the PDR position. Assuming a filament of $12''$
centered at the PDR, we estimate that $\sim10$\% of the \chhhcho{}
emission detected at the dense core corresponds to beam pick-up from
the PDR due to the large beam at 93~GHz ($27''$). The remaining
emission towards the core line of sight could arise from the cloud
surface, as was found for CS\citep{goicoechea2006} and
HCO\cite{gerin2009}.

\subsection*{Propyne}
If \chhhcchE{} and \chhhcchA{} are treated as different
species, the derived rotational temperatures at the PDR are $\sim$70~K
and $\sim$53~K for E and A symmetries, respectively. At the dense core, a
rotational temperature of $\sim70$~K is inferred for the E symmetry. A
rotational temperature cannot be inferred for \chhhcchA{} at the dense
core because the two lines that are detected are faint. The two
symmetries are therefore considered as the same species, resulting in
a rotational temperature of $\sim$55~K and a total column density for
the E- and A-type \chhhcch{} of $\sim2\times10^{13}~\pscm$ at both the
PDR and dense core position. The inferred column density at the PDR
position does not change when the E and A symmetries are separated.

\section{Discussion}
\label{sec:discussion}

Ketene and acetaldehyde are thought to form on the surface of dust
grains. Indeed, \chhhcho{} has been proposed as a candidate for the
7.41~$\mu$m absorption feature observed toward high-mass protostars.
Ketene and acetaldehyde are thought to form together on the ices
through C and H atom additions to CO \citep{herbst2009}. The expected
sequence is
\begin{equation}
\mathrm{CO} \xrightarrow{\mathrm{H}}{} \mathrm{HCO}
\xrightarrow{\mathrm{C}}{} \mathrm{HCCO} 
\xrightarrow{\mathrm{H}}{} \chhco \xrightarrow{\mathrm{2H}}{} \chhhcho.
\end{equation}
Ketene can also be formed from reactions between C$_2$H$_2$ and O in
irradiated \hho{}-rich and CO$_2$-rich ices, as shown by recent
laboratory experiments \citep{hudson2013}. Laboratory experiments also
show that reactions between C$_2$H$_4$ and O can produce acetaldehyde
and its isomer, ethylene oxide (CH$_2$OCH$_2$) \citep{ward2011}.
\chhco{} and \chhhcho{} could also be formed in the gas-phase, through
ion-molecule and neutral-neutral reactions. Indeed, gas-phase models
predict abundances that are comparable to those measured in some of
the high-mass protostars observed by \citet{bisschop2007}.

The exact formation path of HCOOH on ices is unclear, though HCOOH
ices have been observed in star-forming regions
\citep{keane2001}. Several formation paths on grain surfaces have been
proposed in the past. It could form from the addition of H and O atoms
to CO or to CO$_2$\citep{tielens1982}. \citet{garrod2006} proposed
that HCOOH could form through reactions between HCO and OH. More
recently, \citet{ioppolo2011} have studied the hydrogenation of the
HO-CO complex in the laboratory and showed it is an efficient
formation route to HCOOH.

\begin{figure*}[t!]
  \centering
  \includegraphics[width=\textwidth]{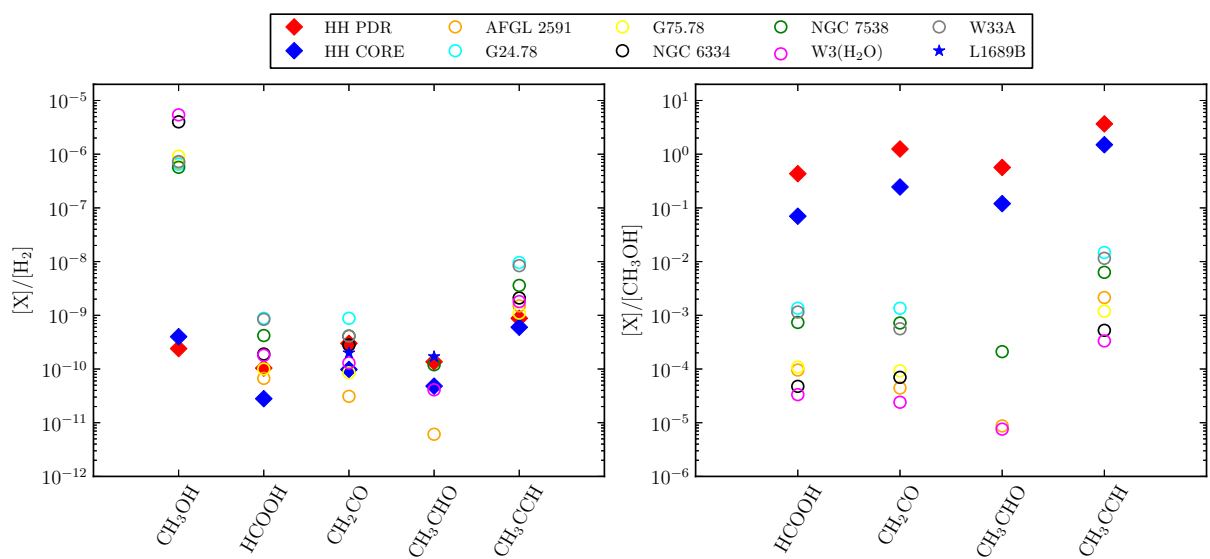} 
  \caption[Abundances with respect to \hh{}]{Abundances with respect
    to \hh{} (\textit{left}) and with respect to \chhhoh{}
    (\textit{right}) toward the hot core sources from
    \citet{bisschop2007} (\textit{open circles}), the cold prestellar
    core from \citet{bacmann2012} (\textit{blue star}) and the
    Horsehead PDR and dense core (\textit{red and blue diamonds}).}
  \label{fig:abund}
\end{figure*}

The abundances derived in the Horsehead PDR for HCOOH, \chhco{}, and
\chhhcho{}, are $3-4$ times larger toward the PDR than toward the
dense core. The case of \chhhcn{} is even more extreme as it is
$\sim30$ times more abundant in the PDR than in the dense
core\citep{gratier2013}. \chhhcch{} is only 1.5 times more abundant in
the PDR than in the dense core. In contrast, methanol is $\sim2$ times
\textit{less} abundant in the PDR than in the dense core. When
comparing the abundances of the different molecules, we found that
\chhhcch{} is one order of magnitude more abundant
($3-4\times10^{-10}$) than \chhco{} and \chhhcho{}
($2-7\times10^{-11}$). Contrary to the other complex molecules, which
present $[\mathrm{X}]/[\chhhoh]$ ratios lower than 1, \chhhcch{} is $\sim4$
times more abundant than methanol in the PDR, and $\sim1.3$ times more
abundant than methanol in the dense core. \citet{oberg2013} also found
large \chhhcch{} abundances ($\sim1$ relative to methanol) toward the
high-mass protostar NGC 7538 IRS9. They found the \chhhcch{}/\chhhoh{}
abundance ratio to be significantly different to what models including
grain surface processes predict, which suggests that an important cold
formation pathway is missing for \chhhcch{}.% However, the observed
%abundances of \chhhcch{} in dark and translucent clouds can be
%reproduced by pure gas-phase models\citep{bisschop2007}.

Fig.~\ref{fig:abund} shows a comparison between the abundances derived
in the Horsehead and those derived toward the hot corino sources from
\citet{bisschop2007} and toward the prestellar core
L1689B\citep{bacmann2012}. \chhhoh{} is several orders of magnitude
more abundant toward the hot core sources than in the Horsehead. The
abundances of the other complex molecules with respect to \hh{} vary
over $\sim1$ order of magnitude between the different hot core
sources, and are comparable to the abundances derived in the
Horsehead. The abundances with respect to \chhhoh{} are also shown in
the right panel of Fig.~\ref{fig:abund}. In this case, the abundances
of complex molecules are $\sim3$ orders of magnitude larger in the
Horsehead than in the hot cores. However, this could be a consequence
of methanol and the other complex molecules tracing different regions
in the hot core sources. Methanol probably traces the hot
($\Tkin>100$~K) gas where species have evaporated from the grains,
while the other complex molecules trace the colder envelope around the
protostars, where the ices have not completely evaporated but can be
photodesorbed.

The fact that we only detect cold molecules (HCOOH, \chhco{},
\chhhcho{} and \chhhcch) and none of the hot molecules (\eg{},
CH$_3$OCH$_3$ and HCOOCH$_3$), is in agreement with the idea that the
cold molecules are zeroth or first generation species formed on the
cold grain surfaces and trace the warm/cold envelope around
protostars, because their formation probably requires little
energy. In the Horsehead, the enhanced abundances toward the PDR
compared to the dense core (for HCOOH, \chhco{}, and \chhhcho{}),
suggests that their formation is more efficient in the presence of
far-UV photons. The similarities between the HCO and \chhhcho{}
emission maps also suggests that the \chhhcho{} abundance at the PDR
could be even higher than estimated here, if the emission arises from
a narrow filament like HCO. This could be the result of an efficient
photodesorption in the PDR, due to the larger radiation field compared
to the dense core, which is consistent with the much lower than unity
E/A ratio (statistical value) we infer from the observations. But it
could also indicate that the formation on the grains itself is more
efficient in the PDR, due to a better mobility of the molecules in ice
mantles. Indeed, recent laboratories experiments have shown that the
diffusion of molecules is active at $\Td\gtrsim30$~K, allowing reactions to
proceed faster when the ices are warmed by far-UV
photons\citep{vinogradoff2013,mispelaer2013}. Dust temperatures in the
Horsehead PDR range from $\Td\sim30$~K in the PDR to $\Td\sim20$~K in
the dense core\citep{goicoechea2009a}.

\section{Conclusions}
\label{sec:conclusions}

We have carried-out an unbiased spectral line survey at 3, 2 and 1mm
with the IRAM-30m telescope in the Horsehead nebula, with an
unprecedented combination of bandwidth, high spectral resolution and
sensitivity. Two positions were observed: the warm photodissociation
region (PDR) and a cold condensation shielded from the UV field,
located less than $40''$ away from the PDR edge. The results of this
survey include 1) the detection of \cfp{}, which can be used as a new
diagnostic of UV illuminated gas and a potential proxy of the \cp{}
emission associated to molecular gas; 2) the detection of a new
species in the ISM, the small hydrocarbon \ccchp{}, which confirms
the top-down scenario of formation of the small hydrocarbons from PAHs
and photo-erosion; 3) the detection of \hhco{}, \chhhoh{} and
\chhhcn{}, which reveals that photo-desorption of ices is an efficient
mechanism to release molecules into the gas phase; 4) and the first
detection of the complex organic molecules, HCOOH, \chhco{},
\chhhcho{} and \chhhcch{} in a PDR, which reveals the degree of
chemical complexity reached in the UV illuminated neutral gas. Complex
molecules are usually considered as hot-core tracers. The detection of
these molecules in PDRs shows that they can survive in the presence of
far-UV radiation, and their formation could even be enhanced due to
the radiation. This opens the possibility of detecting complex
molecules in other far-UV illuminated regions, such as protoplanetary
disks, in the future. From this work we conclude that grain surface
chemistry and non-thermal desorption are crucial processes in the ISM
and therefore must be incorporated into photochemical models to interpret the
observations.

%suggests that
%they can also form in UV-irradiated warm gas. We have shown that
%complex molecules 

\section*{Acknowledgements}
V.G. thanks support from the Chilean Government through the Becas
Chile scholarship program. This work was also funded by grant ANR-09-
BLAN-0231-01 from the French Agence Nationale de la Recherche as part
of the SCHISM project. J.R.G. thanks the Spanish MICINN for funding
support through grants AYA2009-07304 and CSD2009-00038. J.R.G. is
supported by a Ram\'on y Cajal research contract from the Spanish MICINN
and co-financed by the European Social Fund.


%\balance

%If notes are included in your references you can change the title using the following command.

\footnotesize{
\bibliographystyle{rsc}
\bibliography{bib} %your .bib file
}

\begin{appendix}

\section{Observational tables}
\label{app:tables}

\begin{table*}[h!]
\footnotesize{
  \centering
  \caption[Observation parameters of the HCOOH lines]{Observation
    parameters of the deep integrations of the HCOOH lines detected
    toward the PDR and dense core.}
  \begin{threeparttable}
    \begin{tabular}{llrccccccrrr}
      \toprule
      Molecule & Transition & $\nu$ & $E_u$ & $A_{ul}$ & $g_u$ & Line area & Velocity & FWHM &
      $T_{\textrm{peak}}$ & RMS & Peak S/N \\
       & & GHz & K & $\ps$ & & $\mKkms$ & $\kms$ & $\kms$ & mK & mK &\\
      \midrule
      \multicolumn{12}{c}{PDR} \\
      \thcooh{} & $4_{14}-3_{13}$ &  86.546 & 13.6 & $6.0\times10^{-6}$ &  9 & 14.3$\pm$2.5 & 10.63 & 0.41 &  32.6 &  6.0 &   5\\
      \thcooh{} & $4_{04}-3_{03}$ &  89.579 & 10.8 & $7.0\times10^{-6}$ &  9 & 13.9$\pm$2.6 & 10.74 & 0.58 &  22.6 &  5.8 &   4\\
      \thcooh{} & $4_{22}-3_{21}$ &  90.165 & 23.5 & $6.0\times10^{-6}$ &  9 & 10.2$\pm$2.2 & 10.91 & 0.51 &  18.7 &  5.4 &   3\\
      \thcooh{} & $4_{13}-3_{12}$ &  93.098 & 14.4 & $8.0\times10^{-6}$ &  9 & 10.8$\pm$2.4 & 10.54 & 0.51 &  19.7 &  6.0 &   3\\
      \thcooh{} & $5_{15}-4_{14}$ & 108.127 & 18.8 & $1.3\times10^{-5}$ & 11 & 11.3$\pm$2.9 & 10.83 & 0.38 &  27.7 &  9.8 &   3\\
      \thcooh{} & $5_{05}-4_{04}$ & 111.747 & 16.1 & $1.4\times10^{-5}$ & 11 & 13.2$\pm$5.1 & 10.82 & 0.41 &  30.0 &  9.3 &   3\\    
      \thcooh{} & $6_{16}-5_{15}$ & 129.672 & 25.0 & $2.2\times10^{-5}$ & 13 & 17.8$\pm$4.1 & 10.66 & 0.40 &  42.1 & 13.2 &   3\\
      \thcooh{} & $6_{06}-5_{05}$ & 133.767 & 22.5 & $2.5\times10^{-5}$ & 13 & 17.1$\pm$3.9 & 10.57 & 0.36 &  45.2 & 14.1 &   3\\
      \thcooh{} & $6_{24}-5_{23}$ & 135.738 & 35.4 & $2.3\times10^{-5}$ & 13 & 16.1$\pm$3.5 & 10.46 & 0.30 &  49.9 & 13.9 &   4\\
      \midrule
      % Cold Core 
      \multicolumn{12}{c}{CORE} \\
      \thcooh{} & $4_{04}-3_{03}$ &  89.579 & 10.8 & $7.0\times10^{-6}$ &  9 & 17.2$\pm$2.8 & 10.56 & 0.59 &  27.4 &  6.2 &   4\\ 
      \thcooh{} & $4_{13}-3_{12}$ &  93.098 & 14.4 & $8.0\times10^{-6}$ &  9 & 11.4$\pm$2.6 & 10.64 & 0.34 &  31.4 &  6.5 &   5\\ 
      \thcooh{} & $5_{15}-4_{14}$ & 108.127 & 18.8 & $1.3\times10^{-5}$ & 11 & 16.3$\pm$3.5 & 10.61 & 0.38 &  40.1 &  9.9 &   4\\ 
      \thcooh{} & $5_{05}-4_{04}$ & 111.747 & 16.1 & $1.4\times10^{-5}$ & 11 & 11.5$\pm$3.1 & 10.54 & 0.30 &  36.3 &  9.0 &   4\\ 
      \thcooh{} & $6_{06}-5_{05}$ & 133.767 & 22.5 & $2.5\times10^{-5}$ & 13 & 26.2$\pm$5.6 & 10.39 & 0.44 &  55.9 & 13.5 &   4\\ 
      \bottomrule
    \end{tabular}
    \begin{tablenotes}[para,flushleft]
      Note: All temperatures are given in the main beam temperature
      scale. 
    \end{tablenotes}
  \end{threeparttable}
  \label{tab:obs:hcooh}}
\end{table*}

\begin{table*}[h!]
\footnotesize{
  \centering
  \caption[Observation parameters of the $\chhco$ lines]{Observation
    parameters of the deep integrations of the $\chhco$ lines detected
    toward the PDR and dense core.}
  \begin{threeparttable}
    \begin{tabular}{llrrcccccrrr}
      \toprule
      Molecule & Transition & $\nu$ & $E_u$ & $A_{ul}$ & $g_u$ & Line area & Velocity & FWHM &
      $T_{\textrm{peak}}$ & RMS & Peak S/N \\
      & & GHz & K & $\ps$ & & $\mKkms$ & $\kms$ & $\kms$ & mK & mK &\\
      \midrule
      \multicolumn{12}{c}{PDR} \\ 
      \ochhco{} & $4_{13}-3_{12}$  &  81.586 &  22.9 & $5.0\times10^{-6}$ & 27 &  83.1$\pm$10.7 & 10.62 & 0.69 & 113.4 & 19.8 &   6\\ 
      \ochhco{} & $5_{15}-4_{14}$  & 100.095 &  27.5 & $1.0\times10^{-5}$ & 33 &  87.9$\pm$ 3.6 & 10.70 & 0.64 & 128.7 &  7.4 &  17\\ 
      \ochhco{} & $5_{14}-4_{13}$  & 101.981 &  27.8 & $1.1\times10^{-5}$ & 33 &  89.0$\pm$ 3.2 & 10.67 & 0.65 & 127.7 &  7.1 &  18\\ 
      \ochhco{} & $7_{17}-6_{16}$  & 140.127 &  40.0 & $2.9\times10^{-5}$ & 45 &  77.9$\pm$ 4.7 & 10.66 & 0.55 & 133.0 & 12.4 &  11\\ 
      \ochhco{} & $7_{16}-6_{15}$  & 142.769 &  40.5 & $3.1\times10^{-5}$ & 45 &  68.4$\pm$ 8.1 & 10.66 & 0.47 & 135.3 & 25.6 &   5\\ 
      \ochhco{} & $8_{18}-7_{17}$  & 160.142 &  47.6 & $4.5\times10^{-5}$ & 51 & 102.7$\pm$13.5 & 10.74 & 0.72 & 134.2 & 35.7 &   4\\ 
      \ochhco{} & $10_{19}-9_{18}$ & 203.940 &  66.9 & $9.3\times10^{-5}$ & 63 &  45.9$\pm$ 7.0 & 10.72 & 0.67 &  64.8 & 10.6 &   6\\ 
      \pchhco{} & $5_{05}-4_{04}$  & 101.037 &  14.5 & $1.1\times10^{-5}$ & 11 &  45.3$\pm$ 3.1 & 10.81 & 0.70 &  60.8 &  6.8 &   9\\  
      \pchhco{} & $7_{07}-6_{06}$  & 141.438 &  27.2 & $3.1\times10^{-5}$ & 15 &  42.9$\pm$ 6.7 & 10.67 & 0.56 &  71.6 & 17.9 &   4\\ [1ex]
      \ochhco{}$^c$ & $5_{33}-4_{32}$$^a$& 101.002 & 132.8 & $7.0\times10^{-6}$ & 33 &  31.0$\pm$ 4.5 & 10.98 & 0.87 &  33.6 &  8.5 &   4\\ %  
      \ochhco{}$^c$ & $7_{35}-6_{34}$$^b$& 141.402 & 145.4 & $2.5\times10^{-5}$ & 45 &  50.2$\pm$ 8.0 & 10.28 & 1.03 &  45.9 & 16.9 &   3\\ %
      \ochhco{}$^c$ & $7_{25}-6_{24}$  & 141.452 &  79.5 & $2.8\times10^{-5}$ & 15 &  39.1$\pm$ 6.9 & 10.48 & 0.83 &  44.4 & 15.7 &   3\\ %
     \midrule                        
      % Cold Core 
      \multicolumn{12}{c}{CORE} \\
      \ochhco{} & $4_{13}-3_{12}$ &  81.586 & 22.9 & $5.0\times10^{-6}$ & 27 & 34.3$\pm$8.6 & 10.86 & 0.69 &  47.0 & 17.8 &   3\\ 
      \ochhco{} & $5_{15}-4_{14}$ & 100.095 & 27.5 & $1.0\times10^{-5}$ & 33 & 38.9$\pm$3.3 & 10.67 & 0.52 &  70.7 &  7.8 &   9\\ 
      \ochhco{} & $5_{14}-4_{13}$ & 101.981 & 27.8 & $1.1\times10^{-5}$ & 33 & 33.3$\pm$3.2 & 10.65 & 0.47 &  66.6 &  7.8 &   9\\ 
      \ochhco{} & $7_{17}-6_{16}$ & 140.127 & 40.0 & $2.9\times10^{-5}$ & 45 & 30.6$\pm$4.2 & 10.60 & 0.51 &  56.5 & 11.8 &   5\\ 
      \pchhco{} & $5_{05}-4_{04}$ & 101.037 & 14.5 & $1.1\times10^{-5}$ & 11 & 29.5$\pm$3.9 & 10.68 & 0.73 &  37.7 &  8.0 &   5\\ 
      \pchhco{} & $7_{07}-6_{06}$ & 141.438 & 27.2 & $3.1\times10^{-5}$ & 15 & 31.1$\pm$7.0 & 10.68 & 0.69 &  42.5 & 17.0 &   3\\ 
      \bottomrule
    \end{tabular}
    \begin{tablenotes}[para,flushleft]
      Note: All temperatures are given in the main beam
      temperature scale.\\
      $^a$ Blended with the $5_{32}-4_{31}$ line.\\
      $^b$ Blended with the $7_{34}-6_{33}$ line.\\
      $^c$ The line identification is uncertain.
    \end{tablenotes}
  \end{threeparttable}
  \label{tab:obs:ch2co}}
\end{table*}

\begin{table*}[h!]
\footnotesize{
  \centering
  \caption[Observation parameters of the $\chhhcho$ lines]{Observation
    parameters of the deep integrations of the $\chhhcho$ lines
    detected toward the PDR and dense core.}
  \begin{threeparttable}
    \begin{tabular}{llrccccccrrr}
      \toprule
      Molecule & Transition & $\nu$ & $E_u$ & $A_{ul}$ & $g_u$ & Line area & Velocity & FWHM &
      $T_{\textrm{peak}}$ & RMS & Peak S/N \\
      & & GHz  & K & s$^{-1}$ & & $\mKkms$ & $\kms$ & $\kms$ & mK & mK &\\
      \midrule
      \multicolumn{12}{c}{PDR} \\
      \chhhchoE{} & $5_{15}-4_{14}$ &  93.595 & 15.8 & $2.5\times10^{-5}$ & 22 & 57.5$\pm$ 2.5 & 10.75 & 0.63 &  86.0 &  5.2 &  16\\ 
      \chhhchoE{} & $5_{05}-4_{04}$ &  95.947 & 13.9 & $2.8\times10^{-5}$ & 22 & 81.6$\pm$ 3.0 & 10.67 & 0.68 & 113.5 &  6.1 &  19\\ 
      \chhhchoE{} & $5_{14}-4_{13}$ &  98.863 & 16.6 & $3.0\times10^{-5}$ & 22 & 68.2$\pm$ 3.5 & 10.68 & 0.61 & 104.7 &  8.0 &  13\\ 
      \chhhchoE{} & $6_{16}-5_{15}$ & 112.254 & 21.2 & $4.5\times10^{-5}$ & 26 & 41.8$\pm$ 4.1 & 10.59 & 0.47 &  83.4 & 11.1 &   8\\ 
      \chhhchoE{} & $6_{06}-5_{05}$ & 114.940 & 19.4 & $5.2\times10^{-5}$ & 26 & 88.2$\pm$15.3 & 10.65 & 0.67 & 124.3 & 32.0 &   4\\ 
      \chhhchoE{} & $7_{07}-6_{06}$ & 133.831 & 25.9 & $8.2\times10^{-5}$ & 30 & 36.2$\pm$ 7.4 & 10.70 & 0.83 &  41.0 & 16.6 &   2\\ 
      \chhhchoE{} & $7_{16}-6_{15}$ & 138.285 & 28.9 & $8.6\times10^{-5}$ & 30 & 30.6$\pm$ 4.7 & 10.60 & 0.52 &  54.9 & 14.5 &   4\\ 
      \chhhchoE{} & $8_{18}-7_{17}$ & 149.505 & 34.7 & $1.1\times10^{-4}$ & 34 & 20.5$\pm$ 4.2 & 10.81 & 0.33 &  59.1 & 16.8 &   4\\ 
      \chhhchoA{} & $5_{15}-4_{14}$ &  93.581 & 15.7 & $2.5\times10^{-5}$ & 22 & 44.3$\pm$ 2.6 & 10.66 & 0.54 &  76.9 &  6.0 &  13\\ 
      \chhhchoA{} & $5_{05}-4_{04}$ &  95.963 & 13.8 & $2.8\times10^{-5}$ & 22 & 73.1$\pm$ 2.8 & 10.70 & 0.63 & 108.5 &  5.8 &  19\\ 
      \chhhchoA{} & $5_{24}-4_{23}$ &  96.274 & 22.9 & $2.4\times10^{-5}$ & 22 & 13.5$\pm$ 2.3 & 10.55 & 0.48 &  26.7 &  5.6 &   5\\ 
      \chhhchoA{} & $5_{23}-4_{22}$ &  96.633 & 23.0 & $2.4\times10^{-5}$ & 22 & 12.3$\pm$ 1.9 & 10.68 & 0.41 &  28.5 &  5.1 &   6\\ 
      \chhhchoA{} & $5_{14}-4_{13}$ &  98.901 & 16.5 & $3.0\times10^{-5}$ & 22 & 73.8$\pm$ 3.4 & 10.73 & 0.62 & 112.7 &  7.7 &  15\\ 
      \chhhchoA{} & $6_{16}-5_{15}$ & 112.249 & 21.1 & $4.5\times10^{-5}$ & 26 & 56.0$\pm$ 5.1 & 10.68 & 0.69 &  76.3 & 11.4 &   7\\ 
      \chhhchoA{} & $6_{06}-5_{05}$ & 114.960 & 19.4 & $5.0\times10^{-5}$ & 26 & 83.6$\pm$13.6 & 10.71 & 0.67 & 117.5 & 28.8 &   4\\ 
      \chhhchoA{} & $7_{07}-6_{06}$ & 133.854 & 25.8 & $7.9\times10^{-5}$ & 30 & 30.8$\pm$ 3.6 & 10.61 & 0.39 &  73.9 & 11.9 &   6\\ 
      \chhhchoA{} & $7_{25}-6_{24}$ & 135.685 & 35.0 & $7.9\times10^{-5}$ & 30 & 15.4$\pm$ 3.2 & 10.76 & 0.29 &  50.5 & 13.2 &   4\\ 
      \chhhchoA{} & $7_{16}-6_{15}$ & 138.320 & 28.8 & $8.6\times10^{-5}$ & 30 & 52.9$\pm$ 6.8 & 10.74 & 0.49 & 100.6 & 17.0 &   6\\ 
      \midrule
      % Cold Core 
      \multicolumn{12}{c}{CORE} \\
      \chhhchoE{} & $5_{15}-4_{14}$ &  93.595 & 15.8 & $2.5\times10^{-5}$ & 22 & 34.4$\pm$2.2 & 10.71 & 0.74 &  43.5 &  4.2 &  10\\ 
      \chhhchoE{} & $5_{05}-4_{04}$ &  95.947 & 13.9 & $2.8\times10^{-5}$ & 22 & 38.5$\pm$2.4 & 10.66 & 0.63 &  57.1 &  5.4 &  10\\ 
      \chhhchoE{} & $5_{14}-4_{13}$ &  98.863 & 16.6 & $3.0\times10^{-5}$ & 22 & 26.6$\pm$4.0 & 10.63 & 0.60 &  41.5 &  8.8 &   5\\ 
      \chhhchoA{} & $5_{15}-4_{14}$ &  93.581 & 15.7 & $2.5\times10^{-5}$ & 22 & 26.7$\pm$2.0 & 10.62 & 0.57 &  44.1 &  4.6 &  10\\ 
      \chhhchoA{} & $5_{05}-4_{04}$ &  95.963 & 13.8 & $2.8\times10^{-5}$ & 22 & 33.1$\pm$3.1 & 10.67 & 0.80 &  38.9 &  5.7 &   7\\ 
      \chhhchoA{} & $5_{14}-4_{13}$ &  98.901 & 16.5 & $3.0\times10^{-5}$ & 22 & 35.5$\pm$5.7 & 10.74 & 0.74 &  44.8 &  9.5 &   5\\ 
      \chhhchoA{} & $6_{16}-5_{15}$ & 112.249 & 21.1 & $4.5\times10^{-5}$ & 26 & 16.0$\pm$3.5 & 10.62 & 0.55 &  27.2 &  9.9 &   3\\ 
      \bottomrule
    \end{tabular}
    \begin{tablenotes}[para,flushleft]
      Note: All temperatures are given in the main beam temperature scale.
    \end{tablenotes}
  \end{threeparttable}
  \label{tab:obs:ch3cho}}
\end{table*}

\begin{table*}[h!]
\footnotesize{
  \centering
  \caption[Observation parameters of the $\chhhcch$ lines]{Observation
    parameters of the deep integrations of the $\chhhcch$ lines
    detected toward the PDR and dense core.}
  \begin{threeparttable}
    \begin{tabular}{llrccccccrrr}
      \toprule
      Molecule & Transition & $\nu$ & $E_u$ & $A_{ul}$ & $g_u$ & Line area & Velocity & FWHM$^a$ &
      $T_{\textrm{peak}}$ & RMS & Peak S/N \\
      & & GHz & K & s$^{-1}$ & & $\mKkms$ & $\kms$ & $\kms$ & mK & mK &\\
      \midrule
      \multicolumn{9}{c}{PDR} \\
      \chhhcchE{} &  $5_1-4_1$  &  85.456 &  19.5 & $5.93\times10^{-7}$ & 22 & 15.2$\pm$ 2.7 & 10.54 & 0.60 &  23.9 &  6.8 &   3\\ 
      \chhhcchE{} &  $6_1-5_1$  & 102.546 &  24.4 & $1.05\times10^{-6}$ & 26 & 16.0$\pm$ 2.7 & 10.95 & 0.60 &  25.0 &  7.6 &   3\\ 
      \chhhcchE{} &  $8_1-7_1$  & 136.725 &  36.7 & $2.50\times10^{-6}$ & 34 & 16.8$\pm$ 3.7 & 10.53 & 0.60 &  26.3 & 12.2 &   2\\ 
      \chhhcchE{} & $12_1-11_1$ & 205.077 &  71.2 & $8.95\times10^{-6}$ & 50 & 22.3$\pm$ 3.7 & 10.67 & 0.60 &  34.9 &  7.6 &   4\\ 
      \chhhcchE{} & $13_1-12_1$ & 222.163 &  81.9 & $1.14\times10^{-5}$ & 27 & 15.3$\pm$ 3.8 & 10.66 & 0.60 &  23.9 &  7.9 &   3\\ 
      \chhhcchA{} &  $5_0-4_0$  &  85.457 &  12.3 & $6.18\times10^{-7}$ & 22 & 14.6$\pm$ 2.3 & 10.77 & 0.60 &  22.8 &  5.7 &   4\\ 
      \chhhcchA{} &  $6_0-5_0$  & 102.548 &  17.2 & $1.08\times10^{-6}$ & 26 & 10.8$\pm$ 2.7 & 10.72 & 0.60 &  17.0 &  7.1 &   2\\ 
      \chhhcchA{} &  $8_3-7_3$  & 136.705 &  94.6 & $2.25\times10^{-6}$ & 34 & 15.2$\pm$ 3.7 & 10.55 & 0.60 &  23.8 & 11.8 &   2\\ 
      \chhhcchA{} & $12_0-11_0$ & 205.081 &  64.0 & $9.02\times10^{-6}$ & 50 & 22.0$\pm$ 4.0 & 10.76 & 0.60 &  34.4 &  7.9 &   4\\ 
      \chhhcchA{} & $13_0-12_0$ & 222.167 &  74.6 & $1.15\times10^{-5}$ & 27 & 14.1$\pm$ 3.7 & 10.35 & 0.60 &  22.1 &  7.6 &   2\\ 
      \chhhcchA{} & $14_3-13_3$ & 239.211 & 151.1 & $9.96\times10^{-5}$ & 58 & 28.6$\pm$ 5.9 & 10.46 & 0.60 &  44.8 & 12.8 &   3\\ 
      \chhhcchA{} & $14_0-13_0$ & 239.252 &  86.1 & $1.40\times10^{-5}$ & 58 & 54.2$\pm$ 7.5 & 10.63 & 0.60 &  84.9 & 16.1 &   5\\
      \midrule
      % Cold Core 
      \multicolumn{9}{c}{CORE} \\
      \chhhcchE{} & $5_1-4_1$  &  85.456 & 19.5 & $5.93\times10^{-7}$ & 22 & 13.5$\pm$ 1.9 & 10.48 & 0.50 &  25.4 &  5.5 &   4\\ 
      \chhhcchE{} & $6_1-5_1$  & 102.546 & 24.4 & $1.05\times10^{-6}$ & 26 & 20.0$\pm$ 2.9 & 10.48 & 0.50 &  37.6 &  8.9 &   4\\ 
      \chhhcchE{} & $8_1-7_1$  & 136.725 & 36.7 & $2.50\times10^{-6}$ & 34 & 27.4$\pm$ 4.0 & 10.56 & 0.50 &  51.6 & 14.2 &   3\\ 
      \chhhcchA{} & $5_0-4_0$  &  85.457 & 12.3 & $6.18\times10^{-7}$ & 22 & 16.8$\pm$ 2.4 & 10.59 & 0.50 &  31.5 &  6.3 &   4\\ 
      \chhhcchA{} & $6_0-5_0$  & 102.548 & 17.2 & $1.08\times10^{-6}$ & 26 & 25.7$\pm$ 2.8 & 10.74 & 0.50 &  48.2 &  8.5 &   5\\ 
      \bottomrule
    \end{tabular}
    \begin{tablenotes}[para,flushleft]
      Note: All temperatures are given in the main beam temperature
      scale.\\ 
      $^a$ The line width was fixed to guide the Gaussian fit
      at low signal-to-noise ratio.
    \end{tablenotes}
  \end{threeparttable}
  \label{tab:obs:ch3cch}}
\end{table*}

\end{appendix}

\end{document}